\pdfoutput=1
\PassOptionsToPackage{table}{xcolor}
\documentclass[11pt]{article}

\usepackage[]{ACL2023}

\usepackage{times}
\usepackage{latexsym}
\usepackage{listings}
\usepackage{bigdelim}
\usepackage[skip=0cm,list=true,labelfont=it]{subcaption}

\usepackage[T1]{fontenc}

\usepackage[utf8]{inputenc}

\usepackage{microtype}

\usepackage{inconsolata}

\RequirePackage[inline,shortlabels]{enumitem}
\setlist{nosep}

\usepackage{amsmath, amssymb}
\usepackage{graphicx}
\usepackage{booktabs}
\usepackage{multicol, multirow}
\usepackage[ruled,linesnumbered]{algorithm2e}
\usepackage{bbm}
\usepackage{float}
\usepackage{caption}
\usepackage{xcolor}
\usepackage{natbib}
\usepackage{defs}
\usepackage{todonotes}
\usepackage{makecell}
\usepackage{array}
\usepackage{xcolor}
\usepackage{rotating}
\usepackage{soul}
\usepackage{tikz}
\usepackage{multirow}
\usepackage{booktabs}
\usepackage{longtable}

\definecolor{lightyellow}{RGB}{255,249,196}
\definecolor{lightorange}{RGB}{255,224,178}
\definecolor{lightred}{RGB}{255,205,210}
\definecolor{lightpurple}{RGB}{225,190,231}

\SetCommentSty{mycommfont}
\SetKwComment{Comment}{/* }{ */}
\SetKwFor{For}{for (}{) $\lbrace$}{$\rbrace$}
\RestyleAlgo{ruled}

\newcommand{\rawc}{r}
\newcommand{\trawc}{r^T}

\newcommand{\texttosql}{Text-to-SQL}
\newcommand{\NN}{\text{NN}}
\newcommand{\acc}{\text{Acc}}
\newcommand{\hhvy}{\hat{\hvy}}
\newcommand{\simX}{\text{Sim}}
\newcolumntype{R}[1]{>{\raggedleft\arraybackslash}p{#1}}
\newcommand{\pdt}{prod}
\definecolor{green1}{rgb}{0.8,1,0.8}
\definecolor{yellow1}{rgb}{1,1,0.87}
\definecolor{blue1}{rgb}{0.9,0.9,1}
\definecolor{red1}{rgb}{1,0.9,0.9}
\newcommand{\first}[1]{{\cellcolor{green1} #1}}
\newcommand{\second}[1]{{\cellcolor{yellow1} #1}}

\title{Text-to-SQL Calibration: No Need to Ask—Just Rescale Model Probabilities}

\author{Ashwin Ramachandran \\
  UC San Diego \\
  \texttt{ashwinramg@ucsd.edu} \\\And
  Sunita Sarawagi \\
  IIT Bombay \\
  \texttt{sunita.sarawagi@gmail.com} \\}

\begin{document}
\maketitle


\begin{abstract}
Calibration is crucial as large language models (LLMs) are increasingly deployed to convert natural language queries into SQL for commercial databases. In this work, we investigate calibration techniques for assigning confidence to generated SQL queries. We show that a straightforward baseline—deriving confidence from the model’s full-sequence probability—outperforms recent methods that rely on follow-up prompts for self-checking and confidence verbalization. Our comprehensive evaluation, conducted across two widely-used Text-to-SQL benchmarks and multiple LLM architectures, provides valuable insights into the effectiveness of various calibration strategies.
\end{abstract}

\section{Introduction}
As enterprises increasingly leverage large language models (LLMs) to convert natural language queries into SQL programs for their databases, obtaining well-calibrated confidence estimates becomes critical to spot when the generated SQL may be incorrect~\cite{Steyvers2024TheCG,Baan2023UncertaintyIN}. Such calibrated probabilities can help enterprises select among multiple generated SQLs and determine when to defer to a human expert~\cite{wu2024need}.

Following recent study that highlighted RLHF fine-tuned LLMs produce poorly calibrated conditional probabilities~\cite{kadavath2022language,tian-etal-2023-just}, several work have proposed fixes to the calibration of LLM outputs. Most of these propose to ask the LLM with one or more additional questions seeking True/False response, or multiple choice selection, or verbalized confidence~\cite{kadavath2022language,tian-etal-2023-just,ren2023selfevaluation,Zhou2023NavigatingTG,xiong2024can,kapoor2024largelanguagemodelstaught}.  These methods have been primarily evaluated on tasks such as question answering, where the outputs are typically short strings.
 
For the \texttosql\ task we perform an extensive comparison of several recent and traditional calibration techniques 
spanning several closed and open-source LLMs, including LLMs fine-tuned for \texttosql. 
Our study shows that rescaling the model's sequence probability—using a small validation set with well-established techniques like temperature scaling or isotonic regression—yields significantly better calibration than recent prompting-based methods. Among the strategies that rely on model-generated probabilities, \citet{stengel-eskin-van-durme-2023-calibrated} proposed using the minimum token probability to derive sequence-level confidence. In contrast, our study finds that using the full product of token probabilities, which is theoretically better founded too, provides superior calibration, even after identical rescaling.

\section{Related Work}
Calibration of classification models is a classical ML topic~\cite{05predicting,GuoPSW17}, with much work in pre-LLM NLP literature~\cite{kumar2019calibration,desai-durrett-2020-calibration}.  We focus on recent work on calibration of tasks on LLMs .
\paragraph{Calibration of LLMs for short response generation}\citet{kadavath2022language} study LLMs on a variety of tasks and propose to extract confidence by a self-probe using a follow up True/False question to the LLM on whether the generated response was correct.  Probability of True in the follow up question is measured as confidence. \citet{tian-etal-2023-just} further expand the set of prompts asking to verbalize confidence and show that a better strategy for calibration is getting the LLM to generate top-K responses with probabilities. \citet{ren2023selfevaluation} also show that self-evaluation improves calibration. \citet{Zhou2023NavigatingTG} study if language markers like: "I believe", "I am sure.."etc reflect confidence, and show that these language markers do not faithfully reflect uncertainty. \citet{kuhn2023semantic} propose to generate multiple answers with a confidence score assigned by the LLM for each, cluster them based on semantic similarity, and measure entropy over the total confidence across the clusters. \citet{kapoor2024largelanguagemodelstaught} show that the probabilities output by the self-probe methods are not well-calibrated, and proposes to correct the calibration by further fine-tuning.
\citet{xiong2024can} also studies these techniques and additionally introduces PairRank that scores based on the ranking of responses across multiple Top-K sets.
\citet{Xie2024CalibratingLM} proposes to apply adaptive temperature scaling where the temperature is a linear function of the last layer vector.

\nocite{Huang2024CalibratingLG}







\xhdr{Uncertainty for Semantic Parsing and SQL}
\citet{stengel-eskin-van-durme-2023-calibrated} reports lack of calibration of Text-to-SQL systems and measure confidence as the minimum token probability over tokens in the entire predicted SQL sequence.
Another related topics is measuring how well semantic parsing models represent ambiguity in the input by, for example, outputting both ambiguous logical forms in the top-k output~\cite{stengel-eskin2024zero} and~\cite{bhaskar2023}.



\newcommand{\rawcSC}{\rawc_\text{SC}}
\newcommand{\rawcALT}{\rawc_\text{ALT}}
\section{Calibration Methods}\label{sec:whole}

Let $\vx_i$ be an input natural language question on a database schema $\vs$ for which a \texttosql\ model $\cM$ predicted an output SQL $\hvy_i$.  We explore a number of methods of attaching a score $\rawc(\hvy)$ that indicates if  $\hvy_i$ is a correct SQL for $\vx$.  

\xhdr{Pooled Token-level Probabilities} The generative model $\cM$ assigns a probability $P(\hvy|\vx)$ composed out of auto-regressive token probabilities $\Pr(\hat{y_t}|\vx, \hvy_{<t})$.  A natural method is to use these token probabilities for calibration. Let $n$ denote the number of tokens in $\hvy$. These can be converted into a confidence score $\rawc$ for the whole query $\hvy$ by pooling the token probabilities in various ways: 
\begin{enumerate}[leftmargin=0.5cm, itemsep=2pt]
    \item product of probability \(\prod_t^n\Pr(\hat{y_t}|\vx, \hvy_{<t})\) [\textbf{\pdt}] 
    \item geometric mean \(\sqrt[n]{\prod_t^n\Pr(\hat{y_t}|\vx, \hvy_{<t})}\) [\textbf{geo}]
    \item minimum \(\min_{t \in [n]}\Pr(\hat{y_t}|\vx, \hvy_{<t})\) [\textbf{min}]
    \item arithmetic mean \( \frac{1}{n} \sum_{t=1}^n\Pr(\hat{y_t}|\vx, \hvy_{<t})\) [\textbf{avg}]
\end{enumerate}

\xhdr{LLM Self-checks generated SQL} Another emerging trend is asking the LLM to self-reflect on the correctness of the generated SQL. We consider two variants. (1)  \textbf{[Bool]}~\cite{kadavath2022language}  where the LLM is prompted with the context, the predicted SQL and two options (A: SQL is correct, B: SQL is incorrect). We then collect and normalize the probabilities assigned to tokens 'A' and 'B'. The normalized probability of token 'A' is the confidence score $\rawc(\hvy)$.  (2)  \textbf{[Probs]}~\cite{tian-etal-2023-just} Here, given the context and the predicted SQL, the model is asked to estimate the probability that the SQL is correct. This verbalized probability is used for calibration.




\xhdr{Relative score with Variant output SQLs} 
Given the huge difference in the level of difficulty of SQL generation across different questions and schema, it may be difficult to obtain comparable scores across different instances.  Relative scores across alternative SQLs may be more meaningful. Accordingly, we designed this method: First prompt the model $\cM$ to generate multiple structurally diverse SQLs. 
%
%
Denote alternative plausible SQLs $\cY_\vx$ for  $\vx$.  Out of these we eliminate those SQLs that are semantically equivalent to $\hvy$ based on whether $\acc(\hvy,\vy')$ is the same for each $\vy' \in \cY_\vx$.   Then measure the difference in score of the predicted SQL and the best alternative SQL, that is  $\rawcALT=\rawc(\hvy) - \max_{(\hvy') \in \cY_\vx:\acc(\hvy,\hvy')=0} \rawc(\hvy')$. Other measures could be entropy in scores of the alternatives as proposed here~\cite{kuhn2023semantic}.

\section{Experiments}
We compare these calibration methods across different datasets and LLMs. 
\paragraph{Datasets}
We evaluate on 31 database schemas spanning two popular \texttosql\ benchmarks  Spider~\cite{yu2019spider} and BIRD~\cite{li2023llm}  for natural language utterances $\vx$ and their gold SQL $\vy$.  For each of these, we measure calibration of predictions obtained from two different LLMs  GPT-4~\cite{openai2024gpt4}(‘gpt-3.5-turbo-16k’) and CodeS~\cite{li2024codes}.  The prompts used for SQL generation is provided in Table~\ref{tab:prompts-gen}.  Although these models do not guarantee syntactically valid SQL generation, we assume that a DB engine can be easily invoked to check for grammatical validity and eliminate invalid generations. 
The final statistics of our test data appear in Table~\ref{tab:data}.
\begin{table}[ht]
    \centering
    \small
    \begin{tabular}{|l|r|r|r|r|} \hline
        & \multicolumn{2}{|c|}{Spider}  & \multicolumn{2}{c|}{BIRD} \\ 
        & GPT4 & CodeS & GPT4 & CodeS\\ \hline \hline
        Total Queries & \multicolumn{2}{c}{1034} & \multicolumn{2}{|c|}{1534}\\ \hline
        \# databases & \multicolumn{2}{|c|}{20} & \multicolumn{2}{c|}{11} \\ \hline
        \% Correct & 77.6 \% & 59.1 \% & 43.1 \% & 19.6 \% \\
    \hline
    \end{tabular}
    \caption{Summary of datasets used for calibration.}
    \label{tab:data}
\end{table}

\begin{table*}[ht]
    \centering
    \fontsize{8}{9}\selectfont
    \addtolength{\tabcolsep}{-0.5em}
    \def\arraystretch{1.0}
    \begin{tabular}{p{1.9cm} p{0.4cm} p{1.5cm} | R{1.1cm} R{1.1cm} R{1.4cm} R{1.4cm} | R{1.1cm} R{1.1cm} R{1.4cm} R{1.4cm}}
        \toprule
        \multicolumn{3}{c|}{\multirow{2}{*}{\textbf{Method}}} & \multicolumn{4}{|c|}{\textbf{Spider}} & \multicolumn{4}{c}{\textbf{BIRD}}\\
        \cline{4-11}
        & & & \textbf{BS-I$\downarrow$} & \textbf{AUC$\uparrow$} & \textbf{ECE-P$\downarrow$} & \textbf{ECE-I$\downarrow$} & \textbf{BS-I$\downarrow$} & \textbf{AUC$\uparrow$} & \textbf{ECE-P$\downarrow$} & \textbf{ECE-I$\downarrow$} \\
        \midrule \midrule
        \multirow{4}{2cm}{\fontsize{9}{11}\selectfont \textbf{Pooled token-level}\\(Llama 8B)} & \multirow{4}{*}{$\left\{\rule{0mm}{7mm}\right.$} & min & \second{22.0} & 68.9 & 12.0 & 13.0 & 20.6 & 66.7 & 07.3 & 07.6 \\
        && avg & 23.0 & 62.0 & 13.5 & 14.1 & \second{19.7} & \second{71.7} & 06.7 & 07.6 \\
        && \pdt & \first{18.5} & \first{78.5} & \first{10.2} & \second{11.0} & \first{18.9} & \first{74.6} & 08.3 & 07.7 \\
        && geo & 22.9 & 62.0 & 13.2 & 13.6 & 20.1 & 69.7 & \second{06.4} & \first{06.3}\\
        \hline
        \addlinespace
        \multicolumn{3}{l|}{\parbox{4.5cm}{\fontsize{9}{11}\selectfont\textbf{Self-check Bool} (Llama 8B)}} & 23.7 & \second{48.9} & 13.1 & 14.1 &  22.2 & 58.8 & 06.9 & 09.1\\
        \midrule \midrule
        \multirow{4}{2cm}{\fontsize{9}{11}\selectfont \textbf{Pooled token-level}\\(Llama 70B)} & \multirow{4}{*}{$\left\{\rule{0mm}{7mm}\right.$} & min & 22.1 & 65.5 & 13.1 & 14.1 & 20.8 & 65.5 & 07.8 & 08.1 \\
        && avg & 23.4 & 60.3 & 13.6 & 14.8 & 20.0 & 70.8 & \first{06.5} & 07.2 \\
        && \pdt & \second{18.4} & \second{79.7} & \second{11.3} & 11.8 & \second{19.1} & \second{74.7} & 08.1 & 08.5 \\
        && geo & 23.5 & 58.0 & 13.7 & 14.6 & 20.2 & 69.2 & \second{06.7} & 07.0\\
        \hline
        \addlinespace
        \multicolumn{3}{l|}{\parbox{4.5cm}{\fontsize{9}{11}\selectfont\textbf{Self-check Bool} (Llama 70B)}} & \first{17.5} & \first{80.4} & \first{10.6} & \second{11.3} & \first{17.4} & \first{79.6} & 08.0 & 07.2\\
        \hline \hline
        \addlinespace
        \multirow{4}{2cm}{\fontsize{9}{11}\selectfont \textbf{Pooled token-level}\\(CodeS)} & \multirow{4}{*}{$\left\{\rule{0mm}{7mm}\right.$} & min & 22.3 & 63.6 & 12.0 & 12.4 & 20.6 & 65.1 & 07.0 & 06.6 \\
        && avg & 23.5 & 54.1 & 13.5 & 14.2 & 21.5 & 61.8 & 06.5 & 07.2 \\
        && \pdt & \second{19.3} & 74.6 & 11.2 & 11.2 & \second{19.3} & \second{73.0} & 08.2 & 08.0 \\
        && geo & 21.2 & 53.9 & 13.0 & 14.0 & 21.3 & 63.1 & 06.9 & \second{06.3} \\        
        \hline
        \addlinespace
        \multirow{4}{2cm}{\fontsize{9}{11}\selectfont \textbf{Pooled token-level}\\(Codestral)} & \multirow{4}{*}{$\left\{\rule{0mm}{7mm}\right.$} & min & 22.1 & 66.3 & 11.4 & 13.2 & 20.0 & 67.0 & 07.6 & 06.4 \\
        && avg & 22.4 & 60.6 & 12.6 & 13.1 & 19.9 & 70.5 & 06.7 & \first{06.2} \\
        && \pdt & \first{17.4} & \first{78.8} & \second{09.8} & \first{09.6} & \first{18.4} & \first{75.7} & 07.5 & 07.7 \\
        && geo & 22.5 & 59.8 & 13.3 & 13.4 & 20.2 & 69.4 & 07.9 & 06.9 \\
        \hline
        \addlinespace
        \multicolumn{3}{l|}{\parbox{4.5cm}{\fontsize{9}{11}\selectfont\textbf{Self-check Bool} (GPT-4)}} & 20.6 & 70.1 & 12.0 & 11.9 & 19.7 & 70.7 & 07.1 & \first{06.2} \\
        \hline
        \addlinespace
        \multicolumn{3}{l|}{\parbox{4.5cm}{\fontsize{9}{11}\selectfont\textbf{Self-check Bool} (CodeLlama)}} & 23.8 & 60.0 & 13.2 & 14.8 & 22.1 & 62.1 & 09.0 & 10.1 \\
        \hline
        \addlinespace
        \multicolumn{3}{l|}{\parbox{4.5cm}{\fontsize{9}{11}\selectfont\textbf{Self-check Probs} (GPT-4)}} & 22.0 & 59.8 & 13.0 & 13.0 & 21.4 & 58.4 & \second{06.3} & 06.5 \\
        \hline
        \addlinespace
        \multicolumn{3}{l|}{\parbox{4.5cm}{\fontsize{9}{11}\selectfont\textbf{Variant SQLs (Prod)} (CodeS)}} & 19.5 & \second{74.7} & 11.0 & 11.1 & 19.4 & 70.1 & 09.4 & 06.4 \\
        \hline \hline
    \end{tabular}
\caption{The table presents evaluation metrics on the Spider and BIRD datasets. The table has three parts: the first two rows are using the Llama 8B model, the next two rows with Llama 70B and subsequent six rows feature different models. The metrics include Isotonic-scaled Brier score \textbf{(BS-I)}, area under the ROC curve \textbf{(AUC)}, Platt-scaled expected calibration error \textbf{(ECE-P)} and Isotonic-scaled ECE \textbf{(ECE-I)}. Uniform binning is used to calculate ECE-P and ECE-I.  In each section of the table, the best and second best methods are highlighted in green and yellow, respectively for BS-I, AUC and across both ECE-P and ECE-I.}
\label{tab:main-table}
\end{table*}

\begin{figure*} 
    \centering
    \vspace{-4mm} 
    \subfloat{\includegraphics[width=0.9\textwidth]{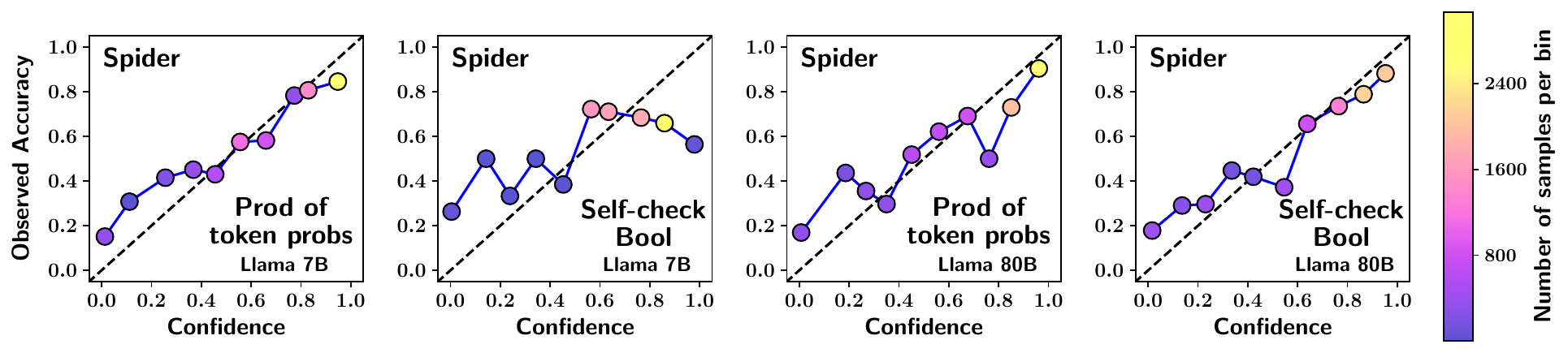}}\\
    \subfloat{\includegraphics[width=0.9\textwidth]{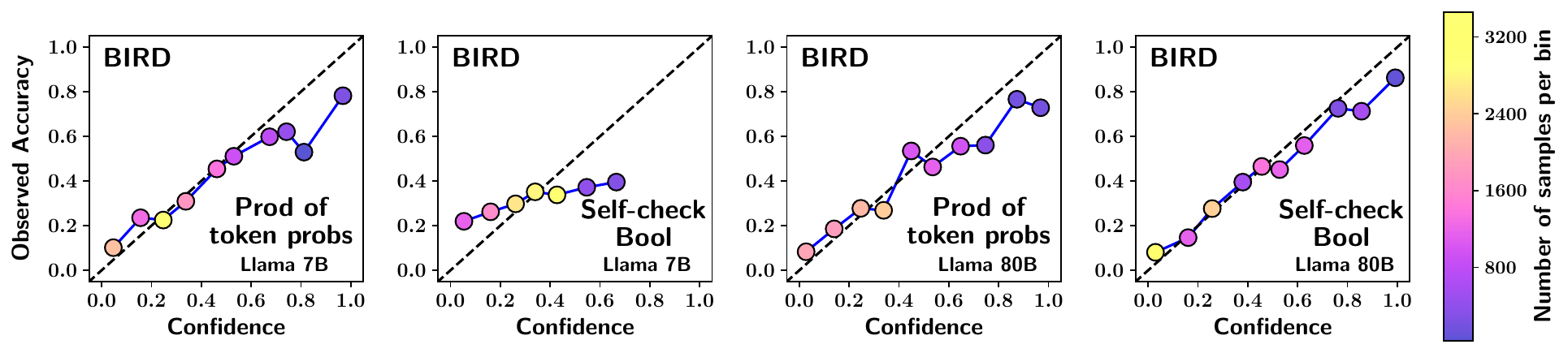}}\vspace{-2mm}
    \caption{The reliability plots compare calibration across different methods and models. The top four plots use predictions from the Spider dataset and bottom four from the BIRD dataset. Plots have been generated with uniform binning and isotonic scaling. A well-calibrated plot aligns closely with the x=y line. Each point is color-coded based on the number of samples in the bin, as indicated by the colorbar on the right.}
    \label{fig:plot-1}
\end{figure*}

%
We compare the different methods using their reliability plots and several calibration measures, as detailed below.  We define these, along with the evaluation protocol in the Appendix~\ref{app:rescale}.

\section{Results}
We report the results across different methods and models in Table~\ref{tab:main-table} and also show the reliability plots in Figure~\ref{fig:plot-1}. Calibration metrics and reliability plots for experiments using other models and monotonic binning have been deferred to the Appendix ~\ref{sec:appendix-whole-query}.  Across these different studies we make two important conclusions:  

\begin{table*}[ht!]
    \centering
    \fontsize{8}{9}\selectfont
    \addtolength{\tabcolsep}{-0.5em}
    \def\arraystretch{1.0}
    \begin{tabular}{p{2.8cm} | R{0.8cm} R{0.8cm} R{0.8cm} | R{0.8cm} R{0.8cm} R{0.8cm} | R{0.8cm} R{0.8cm} R{0.8cm} | R{0.8cm} R{0.8cm} R{0.8cm}}
        \toprule
        \multirow{3}{*}{\textbf{Method}} & \multicolumn{6}{|c|}{\textbf{Spider}} & \multicolumn{6}{c}{\textbf{BIRD}}\\
        \cline{2-13}
        \addlinespace
        & \multicolumn{3}{|c|}{\textbf{Schema-Level}} & \multicolumn{3}{c}{\textbf{Schema-Disjoint}} & \multicolumn{3}{|c|}{\textbf{Schema-Level}} & \multicolumn{3}{c}{\textbf{Schema-Disjoint}}\\
        \cline{2-13}
        \addlinespace
        & \textbf{P$\uparrow$} & \textbf{R$\uparrow$} & \textbf{F1$\uparrow$} & \textbf{P$\uparrow$} & \textbf{R$\uparrow$} & \textbf{F1$\uparrow$} & \textbf{P$\uparrow$} & \textbf{R$\uparrow$} & \textbf{F1$\uparrow$} & \textbf{P$\uparrow$} & \textbf{R$\uparrow$} & \textbf{F1$\uparrow$}\\
        \midrule \midrule
        \addlinespace
        \fontsize{9}{11}\selectfont \textbf{Token-level (Prod)} & 86.1 & 57.3 & 66.9 & 90.4 & 41.0 & 56.5 & 65.1 & 08.7 & 15.1 & 72.8 & 02.8 & 05.4 \\
        \hline
        \addlinespace
        \fontsize{9}{11}\selectfont \textbf{Self-Check Bool} & 85.3 & 54.2 & 63.8 & 88.2 & 32.7 & 47.8 & 64.5 & 21.7 & 20.3 & 86.3 & 01.8 & 03.5\\
        \hline \hline
    \end{tabular}
\caption{The table presents evaluation metrics for the  Token-level (prod) and Self-check Bool methods on the Spider and BIRD dataset, comparing the schema-level and schema-disjoint approaches for a threshold value of 0.9. The metrics include Precision \textbf{(P)}, Recall \textbf{(R)} and F1-score \textbf{(F1)}.}
\label{tab:table-applic}
\end{table*}
\paragraph{Model's own sequence probability (\textbf{prod}) performs better in smaller models and is comparable to self-check method in larger models} 
Recent calibration studies~\cite{tian-etal-2023-just} have found self-check methods to be better, and that could be because they deal with short answers.
~\citet{tian-etal-2023-just} find calibration results differ when using language models fine-tuned with reinforcement learning from human feedback. We use Llama, 7B and 80B, which have undergone several rounds of alignment via supervised fine-tuning (SFT), rejection sampling (RS), and direct preference optimization (DPO) (\citet{llama3.1}). We observe that, even with Llama 70B model, the "prod" pooled token-level approach provides calibration as good as self-check bool. In Llama 8B model, "prod" method performs significantly better.

Further, recent research~\cite{west2024the} highlights the gap between the generative and reasoning capabilities in large language models. From our experiments, we observe the gap becomes more pronounced as the model's parameter count decreases. The product of token probabilities, which is the likelihood of a sequence, serves as a measure of its generative capability, contrasting with self-check which is an assessment of the model's understanding.

\paragraph{\textbf{Prod} aggregation is the best pooled token-level method} \citet{stengel-eskin-van-durme-2023-calibrated} investigate \textbf{min} and \textbf{avg} methods for calibration. We also consider \textbf{prod} aggregation, which represents a more natural choice as it denotes the probability of generating the whole sequence in an auto-regressive model. Our experiments show that \textbf{prod} method yields the best calibration performance, followed by \textbf{min}, with \textbf{avg} and \textbf{geo} providing poor calibration. These findings corroborate early findings~\cite{stengel-eskin-van-durme-2023-calibrated} favoring \textbf{min} aggregation over \textbf{mean}, with our experiments highlighting \textbf{prod} as a significantly better alternative than \textbf{min}.

\subsection*{Schema-level Calibration}
Industries often apply a threshold on calibrated scores to determine when generated SQL can be safely executed without human verification. In our experiments with different thresholds, we observe significant variation in calibration across schema. In our experiments so far, we have used a validation set comprising schemas that were disjoint from the 31 test schemas; we refer to this as the schema-disjoint method.  An interesting question is whether calibration scores can be significantly improved by training the calibration model on labeled data from the same schema being tested; we refer to this as the schema-level method. In Table~\ref{tab:table-applic}, we compare Precision, Recall and F1 scores across varying thresholds for both schema-level and schema-disjoint methods. Our results show that the schema-level method performs better than the schema-disjoint method. We elaborate further in Appendix~\ref{app:schema-level}.

\section{Conclusion}
We study calibration of generated  SQL for LLM based \texttosql\ generation. We find that models show strong calibration when assigning probabilities to queries, that either outperforms or performs as good as verbalization methods. Using product aggregation  for calibration-model assigned probability to the query-provides stronger calibration compared to other methods including minimum aggregation, which was proposed as better by earlier works. This study's insights into model calibration for Text-to-SQL generation can be extended to broader applications such as Python or C++ code generation completion tasks. 

\newpage
\section{Limitations}
Our results are based on the specific models employed in our experiments. Although, we have attempted to ensure the validity of our findings by utilizing different models for each method, we cannot guarantee these results will generalize to other models. This limitation is due to the lack of detailed technical information such as training methodologies for many of the models used. 

Additionally, this limitation restricts our ability to fully explain why the calibration of self-check Bool method is weaker compared to the prod pooled token-level method. Furthermore, our study is restricted to identifying the best calibration methods for generated SQLs, particularly those whose complexity is similar to the SQLs found in the Spider and BIRD datasets.

One potential risk associated with our research is the imperfection of the calibration process. Due to this, the model cannot be applied directly in real world applications with absolute accuracy. The confidence scores predicted by the model should be taken as preliminary assessments. Hence, human evaluation is necessary after the models flag certain instances, ensuring a more reliable decision.

\bibliography{refsCal,refs,anthology,parsing,pubs}

\begin{thebibliography}{33}
\expandafter\ifx\csname natexlab\endcsname\relax\def\natexlab#1{#1}\fi

\bibitem[{AI(2024)}]{llama3.1}
Meta AI. 2024.
\newblock Introducing llama 3.1: Our most capable models to date.
\newblock \url{https://ai.meta.com/blog/meta-llama-3-1/}.
\newblock Accessed: July 23, 2024.

\bibitem[{Baan et~al.(2023)Baan, Daheim, Ilia, Ulmer, Li, Fern{\'a}ndez, Plank, Sennrich, Zerva, and Aziz}]{Baan2023UncertaintyIN}
Joris Baan, Nico Daheim, Evgenia Ilia, Dennis Ulmer, Haau-Sing Li, R.~Fern{\'a}ndez, Barbara Plank, Rico Sennrich, Chrysoula Zerva, and Wilker Aziz. 2023.
\newblock \href {https://api.semanticscholar.org/CorpusID:260316110} {Uncertainty in natural language generation: From theory to applications}.
\newblock \emph{ArXiv}, abs/2307.15703.

\bibitem[{Bhaskar et~al.(2023)Bhaskar, Tomar, Sathe, and Sarawagi}]{bhaskar2023}
Adithya Bhaskar, Tushar Tomar, Ashutosh Sathe, and Sunita Sarawagi. 2023.
\newblock \href {https://openreview.net/forum?id=a0yFO9gKc5} {Benchmarking and improving text-to-{SQL} generation under ambiguity}.
\newblock In \emph{The 2023 Conference on Empirical Methods in Natural Language Processing}.

\bibitem[{{Brier}(1950)}]{brier}
Glenn~W. {Brier}. 1950.
\newblock \href {https://doi.org/10.1175/1520-0493(1950)078<0001:VOFEIT>2.0.CO;2} {{Verification of Forecasts Expressed in Terms of Probability}}.
\newblock \emph{Monthly Weather Review}, 78(1):1.

\bibitem[{Desai and Durrett(2020)}]{desai-durrett-2020-calibration}
Shrey Desai and Greg Durrett. 2020.
\newblock \href {https://doi.org/10.18653/v1/2020.emnlp-main.21} {Calibration of pre-trained transformers}.
\newblock In \emph{Proceedings of the 2020 Conference on Empirical Methods in Natural Language Processing (EMNLP)}, pages 295--302, Online. Association for Computational Linguistics.

\bibitem[{Employees(2024)}]{openai2024gpt4}
OpenAI Employees. 2024.
\newblock \href {http://arxiv.org/abs/2303.08774} {Gpt-4 technical report}.

\bibitem[{Geifman and El-Yaniv(2017)}]{auc}
Yonatan Geifman and Ran El-Yaniv. 2017.
\newblock \href {http://arxiv.org/abs/1705.08500} {Selective classification for deep neural networks}.

\bibitem[{Guo et~al.(2017{\natexlab{a}})Guo, Pleiss, Sun, and Weinberger}]{GuoPSW17}
Chuan Guo, Geoff Pleiss, Yu~Sun, and Kilian~Q. Weinberger. 2017{\natexlab{a}}.
\newblock On calibration of modern neural networks.
\newblock In \emph{Proceedings of the 34th International Conference on Machine Learning, {ICML} 2017, Sydney, NSW, Australia, 6-11 August 2017}, pages 1321--1330.

\bibitem[{Guo et~al.(2017{\natexlab{b}})Guo, Pleiss, Sun, and Weinberger}]{ece}
Chuan Guo, Geoff Pleiss, Yu~Sun, and Kilian~Q. Weinberger. 2017{\natexlab{b}}.
\newblock \href {https://proceedings.mlr.press/v70/guo17a.html} {On calibration of modern neural networks}.
\newblock In \emph{Proceedings of the 34th International Conference on Machine Learning}, volume~70 of \emph{Proceedings of Machine Learning Research}, pages 1321--1330. PMLR.

\bibitem[{Huang et~al.(2024)Huang, Liu, Thirukovalluru, Cohan, and Dhingra}]{Huang2024CalibratingLG}
Yukun Huang, Yixin Liu, Raghuveer Thirukovalluru, Arman Cohan, and Bhuwan Dhingra. 2024.
\newblock \href {https://api.semanticscholar.org/CorpusID:267617073} {Calibrating long-form generations from large language models}.
\newblock \emph{ArXiv}, abs/2402.06544.

\bibitem[{Kadavath et~al.(2022)Kadavath, Conerly, Askell, Henighan, Drain, Perez, Schiefer, Hatfield-Dodds, DasSarma, Tran-Johnson, Johnston, El-Showk, Jones, Elhage, Hume, Chen, Bai, Bowman, Fort, Ganguli, Hernandez, Jacobson, Kernion, Kravec, Lovitt, Ndousse, Olsson, Ringer, Amodei, Brown, Clark, Joseph, Mann, McCandlish, Olah, and Kaplan}]{kadavath2022language}
Saurav Kadavath, Tom Conerly, Amanda Askell, Tom Henighan, Dawn Drain, Ethan Perez, Nicholas Schiefer, Zac Hatfield-Dodds, Nova DasSarma, Eli Tran-Johnson, Scott Johnston, Sheer El-Showk, Andy Jones, Nelson Elhage, Tristan Hume, Anna Chen, Yuntao Bai, Sam Bowman, Stanislav Fort, Deep Ganguli, Danny Hernandez, Josh Jacobson, Jackson Kernion, Shauna Kravec, Liane Lovitt, Kamal Ndousse, Catherine Olsson, Sam Ringer, Dario Amodei, Tom Brown, Jack Clark, Nicholas Joseph, Ben Mann, Sam McCandlish, Chris Olah, and Jared Kaplan. 2022.
\newblock \href {http://arxiv.org/abs/2207.05221} {Language models (mostly) know what they know}.

\bibitem[{Kapoor et~al.(2024)Kapoor, Gruver, Roberts, Collins, Pal, Bhatt, Weller, Dooley, Goldblum, and Wilson}]{kapoor2024largelanguagemodelstaught}
Sanyam Kapoor, Nate Gruver, Manley Roberts, Katherine Collins, Arka Pal, Umang Bhatt, Adrian Weller, Samuel Dooley, Micah Goldblum, and Andrew~Gordon Wilson. 2024.
\newblock \href {http://arxiv.org/abs/2406.08391} {Large language models must be taught to know what they don't know}.

\bibitem[{Kuhn et~al.(2023)Kuhn, Gal, and Farquhar}]{kuhn2023semantic}
Lorenz Kuhn, Yarin Gal, and Sebastian Farquhar. 2023.
\newblock \href {https://openreview.net/forum?id=VD-AYtP0dve} {Semantic uncertainty: Linguistic invariances for uncertainty estimation in natural language generation}.
\newblock In \emph{The Eleventh International Conference on Learning Representations}.

\bibitem[{Kumar and Sarawagi(2019)}]{kumar2019calibration}
Aviral Kumar and Sunita Sarawagi. 2019.
\newblock \href {http://arxiv.org/abs/1903.00802} {Calibration of encoder decoder models for neural machine translation}.

\bibitem[{Li et~al.(2024)Li, Zhang, Liu, Fan, Zhang, Zhu, Wei, Pan, Li, and Chen}]{li2024codes}
Haoyang Li, Jing Zhang, Hanbing Liu, Ju~Fan, Xiaokang Zhang, Jun Zhu, Renjie Wei, Hongyan Pan, Cuiping Li, and Hong Chen. 2024.
\newblock \href {http://arxiv.org/abs/2402.16347} {Codes: Towards building open-source language models for text-to-sql}.

\bibitem[{Li et~al.(2023)Li, Hui, Qu, Yang, Li, Li, Wang, Qin, Cao, Geng, Huo, Zhou, Ma, Li, Chang, Huang, Cheng, and Li}]{li2023llm}
Jinyang Li, Binyuan Hui, Ge~Qu, Jiaxi Yang, Binhua Li, Bowen Li, Bailin Wang, Bowen Qin, Rongyu Cao, Ruiying Geng, Nan Huo, Xuanhe Zhou, Chenhao Ma, Guoliang Li, Kevin C.~C. Chang, Fei Huang, Reynold Cheng, and Yongbin Li. 2023.
\newblock \href {http://arxiv.org/abs/2305.03111} {Can llm already serve as a database interface? a big bench for large-scale database grounded text-to-sqls}.

\bibitem[{Matsubara et~al.(2023)Matsubara, Tax, Mudd, and Guy}]{matsubara2023tce}
Takuo Matsubara, Niek Tax, Richard Mudd, and Ido Guy. 2023.
\newblock \href {http://arxiv.org/abs/2306.14343} {Tce: A test-based approach to measuring calibration error}.

\bibitem[{Niculescu-Mizil and Caruana(2005)}]{05predicting}
Alexandru Niculescu-Mizil and Rich Caruana. 2005.
\newblock Predicting good probabilities with supervised learning.
\newblock In \emph{ICML}.

\bibitem[{Platt(2000)}]{platt}
John Platt. 2000.
\newblock Probabilistic outputs for support vector machines and comparisons to regularized likelihood methods.
\newblock \emph{Adv. Large Margin Classif.}, 10.

\bibitem[{Ren et~al.(2023)Ren, Zhao, Vu, Liu, and Lakshminarayanan}]{ren2023selfevaluation}
Jie Ren, Yao Zhao, Tu~Vu, Peter~J. Liu, and Balaji Lakshminarayanan. 2023.
\newblock \href {http://arxiv.org/abs/2312.09300} {Self-evaluation improves selective generation in large language models}.

\bibitem[{Rozière et~al.(2024)Rozière, Gehring, Gloeckle, Sootla, Gat, Tan, Adi, Liu, Sauvestre, Remez, Rapin, Kozhevnikov, Evtimov, Bitton, Bhatt, Ferrer, Grattafiori, Xiong, Défossez, Copet, Azhar, Touvron, Martin, Usunier, Scialom, and Synnaeve}]{rozière2024code}
Baptiste Rozière, Jonas Gehring, Fabian Gloeckle, Sten Sootla, Itai Gat, Xiaoqing~Ellen Tan, Yossi Adi, Jingyu Liu, Romain Sauvestre, Tal Remez, Jérémy Rapin, Artyom Kozhevnikov, Ivan Evtimov, Joanna Bitton, Manish Bhatt, Cristian~Canton Ferrer, Aaron Grattafiori, Wenhan Xiong, Alexandre Défossez, Jade Copet, Faisal Azhar, Hugo Touvron, Louis Martin, Nicolas Usunier, Thomas Scialom, and Gabriel Synnaeve. 2024.
\newblock \href {http://arxiv.org/abs/2308.12950} {Code llama: Open foundation models for code}.

\bibitem[{Stengel-Eskin et~al.(2024)Stengel-Eskin, Rawlins, and Durme}]{stengel-eskin2024zero}
Elias Stengel-Eskin, Kyle Rawlins, and Benjamin~Van Durme. 2024.
\newblock \href {https://openreview.net/forum?id=qL9gogRepu} {Zero and few-shot semantic parsing with ambiguous inputs}.
\newblock In \emph{The Twelfth International Conference on Learning Representations}.

\bibitem[{Stengel-Eskin and Van~Durme(2023)}]{stengel-eskin-van-durme-2023-calibrated}
Elias Stengel-Eskin and Benjamin Van~Durme. 2023.
\newblock Calibrated interpretation: Confidence estimation in semantic parsing.
\newblock \emph{Transactions of the Association for Computational Linguistics}, 11.

\bibitem[{Steyvers et~al.(2024)Steyvers, Lemus, Kumar, Belem, Karny, Hu, Mayer, and Smyth}]{Steyvers2024TheCG}
Mark Steyvers, Heliodoro~Tejeda Lemus, Aakriti Kumar, Catarina Belem, Sheer Karny, Xinyue Hu, Lukas Mayer, and Padhraic Smyth. 2024.
\newblock \href {https://api.semanticscholar.org/CorpusID:267211649} {The calibration gap between model and human confidence in large language models}.
\newblock \emph{ArXiv}, abs/2401.13835.

\bibitem[{Tian et~al.(2023)Tian, Mitchell, Zhou, Sharma, Rafailov, Yao, Finn, and Manning}]{tian-etal-2023-just}
Katherine Tian, Eric Mitchell, Allan Zhou, Archit Sharma, Rafael Rafailov, Huaxiu Yao, Chelsea Finn, and Christopher Manning. 2023.
\newblock Just ask for calibration: Strategies for eliciting calibrated confidence scores from language models fine-tuned with human feedback.
\newblock In \emph{Proceedings of the 2023 Conference on Empirical Methods in Natural Language Processing}.

\bibitem[{Wang et~al.(2024)Wang, Ren, Yang, Liang, Bai, Chai, Yan, Zhang, Yin, Sun, and Li}]{wang2024macsql}
Bing Wang, Changyu Ren, Jian Yang, Xinnian Liang, Jiaqi Bai, Linzheng Chai, Zhao Yan, Qian-Wen Zhang, Di~Yin, Xing Sun, and Zhoujun Li. 2024.
\newblock \href {http://arxiv.org/abs/2312.11242} {Mac-sql: A multi-agent collaborative framework for text-to-sql}.

\bibitem[{West et~al.(2024)West, Lu, Dziri, Brahman, Li, Hwang, Jiang, Fisher, Ravichander, Chandu, Newman, Koh, Ettinger, and Choi}]{west2024the}
Peter West, Ximing Lu, Nouha Dziri, Faeze Brahman, Linjie Li, Jena~D. Hwang, Liwei Jiang, Jillian Fisher, Abhilasha Ravichander, Khyathi Chandu, Benjamin Newman, Pang~Wei Koh, Allyson Ettinger, and Yejin Choi. 2024.
\newblock \href {https://openreview.net/forum?id=CF8H8MS5P8} {The generative {AI} paradox: {\textquotedblleft}what it can create, it may not understand{\textquotedblright}}.
\newblock In \emph{The Twelfth International Conference on Learning Representations}.

\bibitem[{Wu et~al.(2024)Wu, Tam, Wu, Lin, Lee, and Chen}]{wu2024need}
Cheng-Kuang Wu, Zhi~Rui Tam, Chao-Chung Wu, Chieh-Yen Lin, Hung-yi Lee, and Yun-Nung Chen. 2024.
\newblock I need help! evaluating llm's ability to ask for users' support: A case study on text-to-sql generation.
\newblock \emph{arXiv preprint arXiv:2407.14767}.

\bibitem[{Xie et~al.(2024)Xie, Chen, Lee, Mitchell, and Finn}]{Xie2024CalibratingLM}
Johnathan Xie, Annie~S. Chen, Yoonho Lee, Eric Mitchell, and Chelsea Finn. 2024.
\newblock \href {https://api.semanticscholar.org/CorpusID:272987064} {Calibrating language models with adaptive temperature scaling}.

\bibitem[{Xiong et~al.(2024)Xiong, Hu, Lu, LI, Fu, He, and Hooi}]{xiong2024can}
Miao Xiong, Zhiyuan Hu, Xinyang Lu, YIFEI LI, Jie Fu, Junxian He, and Bryan Hooi. 2024.
\newblock \href {https://openreview.net/forum?id=gjeQKFxFpZ} {Can {LLM}s express their uncertainty? an empirical evaluation of confidence elicitation in {LLM}s}.
\newblock In \emph{The Twelfth International Conference on Learning Representations}.

\bibitem[{Yu et~al.(2019)Yu, Zhang, Yang, Yasunaga, Wang, Li, Ma, Li, Yao, Roman, Zhang, and Radev}]{yu2019spider}
Tao Yu, Rui Zhang, Kai Yang, Michihiro Yasunaga, Dongxu Wang, Zifan Li, James Ma, Irene Li, Qingning Yao, Shanelle Roman, Zilin Zhang, and Dragomir Radev. 2019.
\newblock \href {http://arxiv.org/abs/1809.08887} {Spider: A large-scale human-labeled dataset for complex and cross-domain semantic parsing and text-to-sql task}.

\bibitem[{Zadrozny and Elkan(2002)}]{isotonic}
Bianca Zadrozny and Charles Elkan. 2002.
\newblock \href {https://doi.org/10.1145/775047.775151} {Transforming classifier scores into accurate multiclass probability estimates}.
\newblock \emph{Proceedings of the ACM SIGKDD International Conference on Knowledge Discovery and Data Mining}.

\bibitem[{Zhou et~al.(2023)Zhou, Jurafsky, and Hashimoto}]{Zhou2023NavigatingTG}
Kaitlyn Zhou, Dan Jurafsky, and Tatsunori Hashimoto. 2023.
\newblock \href {https://api.semanticscholar.org/CorpusID:265150666} {Navigating the grey area: How expressions of uncertainty and overconfidence affect language models}.
\newblock In \emph{Conference on Empirical Methods in Natural Language Processing}.

\end{thebibliography}
\bibliographystyle{acl_natbib}

\newpage

\appendix


\section{License}
The Spider and BIRD datasets are distributed under the Creative Commons Attribution-ShareAlike 4.0 International (CC BY-SA 4.0) license. We used code from the \textbf{codes} github repository\footnote{\url{https://github.com/RUCKBReasoning/codes}} released by~\citet{li2024codes}, which is distributed under the Apache-2.0 license. Additionally, we referred to the prompts and execution evaluation scripts from the MAC-SQL Github repository\footnote{\url{https://github.com/wbbeyourself/MAC-SQL/}} released by~\citet{wang2024macsql}; however it's license could not be found. The CodeS models are distributed under the Apache-2.0 license. We used the CodeLlama model in accordance with the Llama 2 community license agreement\footnote{\url{https://github.com/meta-llama/llama/blob/main/LICENSE}}. The Codestral model was used in compliance with the Mistral AI Non-Production License\footnote{\url{https://mistral.ai/news/mistral-ai-non-production-license-mnpl/}}. For inference with GPT-4, we use the paid OpenAI API. We use the Llama models in accordance with the Llama 3.1 community license agreement\footnote{\url{https://github.com/meta-llama/llama-models/blob/main/models/llama3_1/LICENSE}}.

\section{Software and Hardware}
All experiments were run with Python 3.11.5 and PyTorch 2.0.1. Our experiments did not involve any training, but we used GPUs for inference. We used Nvidia A100 (80 GB) GPUs for this purpose. Each inference run took around 2-3 hours with a batch size of 4-6 depending on the model used in the experiment. We release the code for our experiments under Apache-2.0 license.

\section{Experiment Details}
\subsection{Model Inference Details}
We used the Codestral model in 8-bit mode due to hardware constraints. The  Codestral and Llama 70B models were used in 16-bit precision. The models-Llama, CodeS, Codestral and CodeLlama-were sourced from Hugging Face repositories. Specifically, the models were obtained from the following URLs:
\begin{itemize}
    \item Llama 8B: \url{https://huggingface.co/meta-llama/Llama-3.1-8B-Instruct}
    \item Llama 70B: \url{https://huggingface.co/meta-llama/Llama-3.1-70B-Instruct}
    \item CodeS: \url{https://huggingface.co/seeklhy/codes-7b}
    \item Codestral: \url{https://huggingface.co/bullerwins/Codestral-22B-v0.1-hf}
    \item CodeLlama: \url{https://huggingface.co/codellama/CodeLlama-7b-Instruct-hf}
\end{itemize}
Parameter count for each of the models are presented in table~\ref{tab:paramter-count}. For every model, we use the instruction prompts released by the model developers.

\begin{table}[]
    \centering
    \begin{tabular}{c|c}
        \toprule
        Model & Parameters (in Billions) \\
        \midrule\midrule
        Llama & 8 \& 70 \\\hline
        CodeS & 7 \\\hline
        Codestral & 22\\\hline
        CodeLlama & 7 \\\hline
        GPT-4 & -\\
        \hline\hline
    \end{tabular}
    \caption{Parameters in models used for experiments.}
    \label{tab:paramter-count}
\end{table}
\subsection{Evaluation metrics}
\label{app:rescale}
\paragraph{Metrics}
Each data sample comprises of five parts: $(\vx_i, \vy_i, \hvy_i, a_i, \rawc_i)$ where $\vx_i$ is natural language question, $\vy_i$ is gold SQL, $\hvy_i$ is predicted SQL,  $a_i = \acc(\vy_i,\hvy_i)$ is the 0/1 label indicating whether the predicted SQL $\hvy$ produces identical execution results as the gold SQL $\vy$, disregarding the order of columns or rows in the result set, and 
$\rawc_i$ is the confidence value returned by a method evaluated.  The raw confidence scores returned by most methods often need to be monotonically transformed for recalibration~\cite{05predicting,GuoPSW17}.
%
We consider two options~(1) Platt scaling \textbf{(P)}~\cite{platt} and  (2) Isotonic regression \textbf{(I)}~\cite{isotonic} as described below:

\xhdr{Platt and Isotonic Rescaling}
In Platt scaling the raw score $\rawc_i$ is sigmoid scaled with two parameters temperature $t$ and bias $b$, to maximize the likelihood of given $a_i$ under a model $\sigma(t \rawc_i + b)$ where  $\sigma(z)=\frac{1}{1+e^{-z}}$.  In  Isotonic regression \textbf{(I)}~\cite{isotonic} $\rawc_i$ is adjusted so that $\sum_i (a_i - T(\rawc_i))$ is minimized, where $T$ is a step-wise constant isotonic (non-decreasing) function. 

We denote the calibrated confidence score as $\trawc_i$. We use five-fold cross validation using five schema-disjoint splits. In each fold, one split is used for tuning parameters of calibration while the remaining four splits as test data.

\xhdr{Reliability plots}
The data samples are grouped into several bins based on their confidence scores. For each bin, the average confidence is plotted against the observed accuracy, which is the proportion of samples in the bin with \(a_i=1\). By default, all bins have a fixed size (\textbf{Uniform Binning}). We also try the Constrained Pool Adjacent Violators Algorithm~\cite{matsubara2023tce} \textbf{(Monotonic Binning)} which decides the binning such that the average difference between the average observed accuracy and confidence, weighted by the number of samples in each bin is minimized.

\xhdr{Calibration measures}  We measure the Brier score~\cite{brier} which is the mean square difference between calibrated confidence and correctness: \( \frac{1}{n}\sum_i^n (a_i - \trawc_i)^2\). We report these after Platt and Isotonic calibration \textbf{(BS-P/BS-I)}.
We also assess \textbf{AUC}~\cite{auc} after ranking instances on their attached confidence. Note, since both Platt and Isotonic are monotonic, the AUC ranking is on the raw scores. Finally, we measure the expected calibration error~\cite{ece} 
after Platt and Isotonic calibration \textbf{(ECE-P/ECE-I)}. ECE partitions confidence scores (between 0 and 1) into $B=10$ bins and measures 
the mean absolute difference between the average observed accuracy and average confidence, weighted by the number of samples in each bin: \(\sum_{j=1}^{B} \frac{|B_j|}{n}\left| \frac{1}{|B_j|} \sum_{i \in B_j} (a_i - c_i )\right|\), where $c_i$ is either $\rawc_i$ or $\trawc_i$ and the bins $B$ are determined either using uniform binning or monotonic binning method. 

\xhdr{Evaluation Protocol}
\label{app:eval-protocol}
We ensure the robustness of our results by performing experiments using LLMs with different architectures and training methods. We choose four open-source models, CodeS, Codestral\footnote{\url{https://mistral.ai/news/codestral/}} Llama, 8B and 70B \cite{llama3.1} for pooled token level experiments since we need access to the token probabilities. CodeS is specifically optimized for Text-to-SQL generation tasks. Codestral is designed for code generation tasks and has demonstrated superior performance in Text-to-SQL tasks compared to popular open-source models such as Llama-3\footnote{\url{https://ai.meta.com/blog/meta-llama-3/}} and CodeLlama~\cite{rozière2024code}. Llama models are general purpose instruction following LLMs aligned to human preferences. Given the context and predicted SQL, we collect the probabilities assigned by the models to each token of the predicted SQL.

We utilize GPT-4, CodeLlama, Llama 8B and 70B for our self-check experiments due to their reasoning capabilities and ability in understanding self-check questions.
To generate variant output SQLs, we prompt GPT-4 to produce 10 diverse SQLs given the context. For each generated SQL, we calculate the prod pooled token-level confidence score using CodeS.
The prompts used for the experiments are presented in Table~\ref{tab:prompts}, \ref{tab:prompts-2} and \ref{tab:prompts-3}. Specific inference details are deferred to the Appendix.

\section{Prompts}
We adapt the prompts from ~\cite{li2024codes}, ~\cite{wang2024macsql} and ~\cite{tian-etal-2023-just} for pooled token-level, for self-check Bool and  for self-check Probs experiments respectively. We present the prompts used in experiments in tables~\ref{tab:prompts}, ~\ref{tab:prompts-2} and ~\ref{tab:prompts-3}. The prompts used to generate predictions are presented in ~\ref{tab:prompts-gen}.

\section{Additional Results}
\label{sec:appendix-whole-query}

\paragraph{Per-schema Calibration}
\label{app:schema-level}
In industry applications, thresholds are typically set for calibrated scores to determine when generated SQL can be executed without human evaluation. In our experiments so far, we have used a validation set comprising schemas that were disjoint from the 31 test schemas. We call this approach, schema-disjoint. In Tables~\ref{tab:per-schema-disjoint} and ~\ref{tab:per-schema-disjoint-bird}, we report calibration metrics for each schema using the schema-disjoint method for the token-level prod method with the Llama 70B. We observe significant variation across different schemas.

We evaluate our calibration methods using a schema-level method, where calibration is performed using a validation set sourced from the same schema. Tables~\ref{tab:per-schema} and ~\ref{tab:per-schema-level-bird} present calibration metrics for each schema under the schema-level method for the token-level prod method with the Llama 70B.

Additionally, in Tables~\ref{tab:prod-spider} and ~\ref{tab:prod-bird}, we compare the Precision, Recall and F1 scores for schema-level and schema disjoint approaches, using the Token level prod method with the Llama 70B on the Spider and BIRD dataset, respectively. Tables~\ref{tab:self-spider} and ~\ref{tab:self-bird} show the metrics for the self-check Bool method with the Llama 70B on the Spider and BIRD dataset respectively.

We observe a substantial difference in Recall between the schema-disjoint and schema-level methods, leading to poor F1-scores for schema disjoint approach. These results highlight the limitations of the schema-disjoint method and motivate further exploration of schema-level calibration strategies to improve performance.

\newpage
\begin{table*}[ht]
    \centering
    \begin{tabular}{>{\raggedright\arraybackslash}p{1.5cm}>{\raggedright\arraybackslash}p{12cm}}
        \hline
        Method & Prompt Template \\
        \hline
            Pooled token-level &  \makecell[{{p{15cm}}}]{You are provided with a sqlite database schema and a user question. Your task \\ is to generate a sqlite query which can be executed on the sqlite database.\\
            database schema : \\
            table \{table name\}, columns = [\{table name.column\_name\} (\{data type\} | \\ \{is primary key?\} | values: \{sample values\} ), ...] \\
            ..\\
            foreign keys : \\ \{foreign keys\}\\
            matched contents : None\\
            \{question\}\\
            \{SQL\}}\\
        \hline
        Self check Bool & \makecell[{{p{100cm}}}]{
           [Instruction]\\
            Complete SQL query only and with no explanation.\\\relax
            [Constraints]\\
            - In `SELECT <column>`, just select needed 
            columns in the [Question]\\ without any 
            unnecessary column or value\\\vspace{0.05cm}
            - In `FROM <table>` or `JOIN <table>`, 
            do not include unnecessary table\\\vspace{0.05cm}
           - If use max or min func, `JOIN <table>` FIRST, THEN use \\`SELECT MAX(<column>)` or `SELECT MIN(<column>)`\\\vspace{0.05cm}
           - If [Value examples] of <column> has 'None' or None, use \\`JOIN <table>` or `WHERE <column> is NOT NULL` is better\\\vspace{0.05cm}
           - If using `ORDER BY <column> ASC|DESC`, add `GROUP BY <column>` \\before to select             distinct values\\\relax
            [Query]\\
            -- \{question\} \\\relax
            [Evidence] \\
            \{evidence\}\\\relax
            [Database info]\\
            \# Table: \{table name\}\\\relax
            [\\\relax
            (\{column name\}, \{description of column\}. Value examples: [\{sample values\}].),\\
            ..
            \\\relax
            [Foreign keys]\\
            \{foreign keys\}\\
            The proposed SQL for the query is:\\\relax
            [SQL]\\
            ```sql\\
            \{sql\}\\
            ```\\
            }\\ \hline
    \end{tabular}
    \caption{Prompt templates for the methods, Pooled token-level and Self-check Bool.}
    \label{tab:prompts}
\end{table*}

\begin{table*}[ht]
    \centering
    \begin{tabular}{>{\raggedright\arraybackslash}p{1.5cm}>{\raggedright\arraybackslash}p{12cm}}
        \hline
        Method & Prompt Template \\
        \hline
        Self check Probs & \makecell[{{p{100cm}}}]{
           [Instruction]\\
            Complete SQL query only and with no explanation.\\\relax
            [Constraints]\\
            - In `SELECT <column>`, just select needed 
            columns in the [Question]\\ without any 
            unnecessary column or value\\\vspace{0.05cm}
            - In `FROM <table>` or `JOIN <table>`, 
            do not include unnecessary table\\\vspace{0.05cm}
           - If use max or min func, `JOIN <table>` FIRST, THEN use \\`SELECT MAX(<column>)` or `SELECT MIN(<column>)`\\\vspace{0.05cm}
           - If [Value examples] of <column> has 'None' or None, use \\`JOIN <table>` or `WHERE <column> is NOT NULL` is better\\\vspace{0.05cm}
           - If using `ORDER BY <column> ASC|DESC`, add `GROUP BY <column>` \\before to select             distinct values\\\relax
            [Query]\\
            -- \{question\} \\\relax
            [Evidence] \\
            \{evidence\}\\\relax
            [Database info]\\
            \# Table: \{table name\}\\\relax
            [\\\relax
            (\{column name\}, \{description of column\}. Value examples: [\{sample values\}].),\\
            ..
            \\\relax
            [Foreign keys]\\
            \{foreign keys\}\\
            The proposed SQL for the query is:\\\relax
            [SQL]\\
            ```sql\\
            \{sql\}\\
            ```\\
            Provide your best guess and the probability that it is correct (0.0 to 1.0).\\
            Give ONLY the probability, no other words or explanation.\\
            For example:\\
            Probability: <the probability between 0.0 and 1.0 that your guess is correct,\\ without any extra commentary whatsoever; just the probability!>\\
            }\\ \hline
    \end{tabular}
    \caption{Prompt templates for the method, Self-check Probs.}
    \label{tab:prompts-2}
\end{table*}

\begin{table*}[ht]
    \centering
    \begin{tabular}{>{\raggedright\arraybackslash}p{1.5cm}>{\raggedright\arraybackslash}p{12cm}}
        \hline
        Method & Prompt Template \\
        \hline
        Generation of variant SQLs & \makecell[{{p{100cm}}}]{
            When executing SQL below, some errors
            occurred, please fix up SQL based 
            on \\query and database info.\\
            Solve the task step by step if you 
            need to. \\ Use SQL format in the code 
            block, and indicate script type in the code
            block. \\ When you find an answer, verify the 
            answer carefully. Include \\verifiable 
            evidence in your response if possible.\\\relax
            [Constraints]\\
            - In `SELECT <column>`, just select needed 
            columns in the [Question]\\ without any 
            unnecessary column or value\\\vspace{0.05cm}
            - In `FROM <table>` or `JOIN <table>`, 
            do not include unnecessary table\\\vspace{0.05cm}
           - If use max or min func, `JOIN <table>` FIRST, THEN use \\`SELECT MAX(<column>)` or `SELECT MIN(<column>)`\\\vspace{0.05cm}
           - If [Value examples] of <column> has 'None' or None, use \\`JOIN <table>` or `WHERE <column> is NOT NULL` is better\\\vspace{0.05cm}
           - If using `ORDER BY <column> ASC|DESC`, add `GROUP BY <column>` \\before to select             distinct values\\\relax
            [Query]\\
            -- \{question\} \\\relax
            [Evidence] \\
            \{evidence\}\\\relax
            [Database info]\\
            \# Table: \{table name\}\\\relax
            [\\\relax
            (\{column name\}, \{description of column\}. Value examples: [\{sample values\}].),\\
            ..
            \\\relax
            [Foreign keys]\\
            \{foreign keys\}\\
            Generate ten structurally diverse SQLs for the above query}\\ \hline
    \end{tabular}
    \caption{Prompt templates to generate Variant SQLs.}
    \label{tab:prompts-3}
\end{table*}

\begin{table*}[ht]
    \centering
    \small
    \begin{tabular}{>{\raggedright\arraybackslash}p{1.5cm}
    >{\raggedright\arraybackslash}p{1.5cm}
    >{\raggedright\arraybackslash}p{12cm}}
        \toprule
        \textbf{LLM} & \textbf{Dataset} & \textbf{Prompt Template} \\
        \midrule
        CodeS & BIRD & \makecell[{{p{12cm}}}]{
            You are provided with a sqlite database schema and a user question along with a hint to help create an SQL query.\\ 
            Your task is to generate a sqlite query which can be executed on the sqlite database.\\
            \textbf{Input:}\\
            Schema: \\
            \{schema\}\\
            CREATE TABLE table\_name (\\
            column1 datatype CONSTRAINT constraint\_name1,\\
            ...\\
            CONSTRAINT constraint\_name3 PRIMARY KEY \\(column\_name),\\
            CONSTRAINT constraint\_name6 FOREIGN KEY (column\_name) \\REFERENCES other\_table \\
            ...\\
            ...\\
            );\\
            Question: \\
            \{question\}\\
            Hint: \{evidence\}\\
            \textbf{Output:}\\
            SQL: 
        } \\
        \hline
        \addlinespace
        CodeS & Spider & \makecell[{{p{12cm}}}]{
            You are provided with a sqlite database schema and a user question to help create an SQL query.\\ 
            Your task is to generate a sqlite query which can be executed on the sqlite database.\\
            \textbf{Input:}\\
            Schema: \\
            \{schema\}\\
            CREATE TABLE table\_name (\\
            column1 datatype CONSTRAINT constraint\_name1,\\
            column2 datatype CONSTRAINT constraint\_name2,\\
            ...\\
            CONSTRAINT constraint\_name3 PRIMARY KEY \\(column\_name),\\
            ...\\
            );\\
            Question: \\
            \{question\}\\
            \textbf{Output:}\\
            SQL: 
        } \\
        \hline
        \addlinespace
        GPT-4 & Spider & \makecell[{{p{12cm}}}]{
            \#\#\# Complete sqlite SQL query only and with no explanation\\
            \#\#\# Sqlite SQL tables, with their properties: \\
            \#\\
            \# \{table name\}(\{column names\})\\
            ..\\
            \#\\
            \#\#\# \{question\}\\
            SELECT\\
        } \\
        \hline
        \addlinespace
        GPT-4 & BIRD & \makecell[{{p{12cm}}}]{
            \#\#\# Complete sqlite SQL query only and with no explanation\\
            \#\#\# Sqlite SQL tables, with their properties: \\
            \#\\
            \# \{table name\}(\{column names\})\\
            ..\\
            \#\\
            \# Evidence: \{evidence\}
            \#\#\# \{question\}\\
            SELECT\\
        } \\
        \bottomrule
    \end{tabular}
    \caption{Prompt templates for the generating SQL predictions for Spider and BIRD data using CodeS and GPT4.}
    \label{tab:prompts-gen}
\end{table*}


\begin{table*}[ht]
    \centering
    \fontsize{9}{10}\selectfont
    \addtolength{\tabcolsep}{-0.5em}
    \def\arraystretch{1.0}
    \begin{tabular}{p{0.3cm} p{4.3cm} p{0.7cm} | R{1.1cm} R{1.1cm} R{1.1cm} R{1.1cm} R{1.1cm}}
        \toprule
        \multicolumn{3}{c|}{\textbf{Schema}} & \textbf{P$\uparrow$}& \textbf{R$\uparrow$}& \textbf{F1$\uparrow$}& \textbf{BS-I$\downarrow$} & \textbf{AUC$\uparrow$} \\
        \midrule \midrule
        \addlinespace
        & Employee hire evaluation & & 100.0 & 38.2 & 55.3 & 11.1 & 91.5 \\
        \hline
        \addlinespace
        & Singer & & 96.9 & 48.0 & 64.2&11.8 & 83.1\\
        \hline
        \addlinespace
        & cre Doc Template Mgt & & 99.6 & 42.4 & 59.5 &12.4 & 90.5 \\
        \hline
        \addlinespace
        & Poker Player & & 100.0 & 42.3 & 59.5 & 12.0 & 92.7\\
        \hline
        \addlinespace
        & Museum Visit & &100.0 & 30.4 & 46.6 &12.5 & 92.4 \\
        \hline
        \addlinespace
        & WTA 1 & & 98.6 & 40.1 & 57.0 &12.7 & 92.7 \\
        \hline
        \addlinespace
        & Network 1 & &89.1 & 51.5 & 65.3&13.2 & 83.4 \\
        \hline
        \addlinespace
        & Orchestra & & 100.0 & 31.4 & 47.8&14.1 & 92.0\\
        \hline
        \addlinespace
        & Battle Death & & 90.2 & 61.7 & 73.3 &14.8 & 92.5 \\
        \hline
        \addlinespace
        & Voter 1 & & 100.0 & 55.4 & 71.3 &14.9 & 82.0\\
        \hline
        \addlinespace
        & Concert Singer & &93.2 & 40.1 & 56.0 &16.6 & 76.7\\
        \hline
        \addlinespace
        & TV Show & & 90.5 & 36.8 & 52.3&17.2 & 76.8\\
        \hline
        \addlinespace
        & Flight 2 & & 86.8 & 38.5 & 53.4 & 17.3 & 70.7\\
        \hline
        \addlinespace
        & Real Estate Properties & & 76.5 & 54.2 & 63.4 &20.4 & 58.3 \\
        \hline
        \addlinespace
        & Pets 1 & &100.0 & 22.0 & 36.0 & 21.5 & 74.4 \\
        \hline
        \addlinespace
        & Student Transcript Tracking & & 90.1 & 53.2 & 66.9 &21.9 & 85.0\\
        \hline
        \addlinespace
        & Dog Kennels & & 82.2 & 40.5 & 54.2 &22.3 & 78.4\\
        \hline
        \addlinespace
        & Car 1 & & 90.9 & 38.0 & 53.6 & 22.6 & 84.5\\
        \hline
        \addlinespace
        & Course Teach & & 83.6 & 26.1 & 39.8 &26.2 & 55.5 \\
        \hline
        \addlinespace
        & World 1 & &73.1 & 50.2 & 59.5 &29.2 & 76.5 \\
        \hline \hline
    \end{tabular}
\caption{The table presents metrics for the \textbf{token-level prod} method using the Llama 70B model on the Spider dataset using the schema-disjoint approach for each database. The metrics include Precision \textbf{(P)}, Recall \textbf{(R)}, F1-score \textbf{(F1)}, Isotonic-scaled Brier score (BS-I) and area under the ROC curve (AUC).}
\label{tab:per-schema-disjoint}
\end{table*}

\begin{table*}[ht]
    \centering
    \fontsize{9}{10}\selectfont
    \addtolength{\tabcolsep}{-0.5em}
    \def\arraystretch{1.0}
    \begin{tabular}{p{0.3cm} p{4.3cm} p{0.7cm} | R{1.1cm} R{1.1cm} R{1.1cm} R{1.1cm} R{1.1cm}}
        \toprule
        \multicolumn{3}{c|}{\textbf{Schema}} & \textbf{P$\uparrow$} &\textbf{R$\uparrow$} &\textbf{F1$\uparrow$} &\textbf{BS-I$\downarrow$} & \textbf{AUC$\uparrow$}\\
        \midrule \midrule
        \addlinespace
        & Thrombosis Prediction && 100.0 & 00.0 & 00.0 &12.3 & 81.2  \\
        \hline
        \addlinespace
        & California Schools && 100.0 & 00.7 & 01.5 &12.8 & 83.4\\
        \hline
        \addlinespace
        &Financial&&100.0 & 01.0 & 01.9 & 16.1 & 76.5 \\
        \hline
        \addlinespace
        &Debit Card Specializing&&100.0 & 00.9 & 01.7 &16.5 & 74.5\\ 
        \hline
        \addlinespace
        &Toxicology&&72.7 & 02.6 & 05.0 & 17.9 & 77.4\\
        \hline
        \addlinespace
        &European Football 2&&75.0 & 03.2 & 06.2 &18.1 & 74.2 \\
        \hline
        \addlinespace
        &Student Club&&100.0 & 01.2 & 02.3 &20.4 & 78.4 \\
        \hline
        \addlinespace
        &Card Games&&59.3 & 09.7 & 16.7 &21.0 & 74.0 \\
        \hline
        \addlinespace
        & Codebase Community && 86.7 & 03.8 & 07.2 & 22.1 & 74.8\\ 
        \addlinespace
        \hline
        & Formula 1 && 75.0 & 01.9 & 03.6 & 23.1 & 66.1\\
        \addlinespace
        \hline
        &Superhero&&76.9 & 01.9 & 03.8 &23.9 & 75.3\\       
        \hline \hline
    \end{tabular}
\caption{The table presents metrics for the \textbf{token-level prod} method using the Llama 70B model on the BIRD dataset using the schema-disjoint approach for each database. The metrics include Precision \textbf{(P)}, Recall \textbf{(R)}, F1-score \textbf{(F1)}, Isotonic-scaled Brier score (BS-I) and area under the ROC curve (AUC).}
\label{tab:per-schema-disjoint-bird}
\end{table*}
\begin{table*}[ht]
    \centering
    \fontsize{9}{10}\selectfont
    \addtolength{\tabcolsep}{-0.5em}
    \def\arraystretch{1.0}
    \begin{tabular}{p{0.3cm} p{4.3cm} p{0.7cm} | R{1.1cm} R{1.1cm} R{1.1cm} R{1.3cm} R{1.3cm}}
        \toprule
        \multicolumn{3}{c|}{\textbf{Schema}} & \textbf{P$\uparrow$} & \textbf{R$\uparrow$} & \textbf{F1$\uparrow$} & \textbf{BS-I$\downarrow$} & \textbf{AUC$\uparrow$} \\
        \midrule \midrule
        \addlinespace
        & Employee hire evaluation & & 93.9 & 85.3 & 89.4 & 9.2 & 91.5 \\
        \hline
        \addlinespace
        & Singer & & 87.6 & 76.0 & 81.4 & 16.1 & 83.2\\
        \hline
        \addlinespace
        & cre Doc Template Mgt & & 96.1 & 71.9 & 82.3 & 11.1 & 90.5  \\
        \hline
        \addlinespace
        & Poker Player & & 95.7 & 80.2 & 87.3 & 13.7 & 92.8 \\
        \hline
        \addlinespace
        & Museum Visit & &84.2 & 85.7 & 85.0 & 20.1 & 92.2   \\
        \hline
        \addlinespace
        & WTA 1 & & 96.3 & 64.3 & 77.1 & 10.6 & 92.8  \\
        \hline
        \addlinespace
        & Network 1 & &88.8 & 52.4 & 65.9 & 13.2 & 83.4 \\
        \hline
        \addlinespace
        & Orchestra & & 94.7 & 89.6 & 92.1 & 10.3 & 92.1\\
        \hline
        \addlinespace
        & Battle Death & & 79.1 & 56.7 & 66.0 & 16.3 & 92.6\\
        \hline
        \addlinespace
        & Voter 1 & & 87.2 & 73.9 & 80.0 & 18.2 & 81.5 \\
        \hline
        \addlinespace
        & Concert Singer & & 88.8 & 46.7 & 61.2 & 18.1 & 76.7  \\
        \hline
        \addlinespace
        & TV Show & & 82.1 & 47.8 & 60.4 & 17.7 & 77.0\\
        \hline
        \addlinespace
        & Flight 2 & & 83.5 & 47.8 & 60.8 & 17.9 & 71.1\\
        \hline
        \addlinespace
        & Real Estate Properties & & 66.7 & 55.6 & 60.6 & 33.1 & 58.1 \\
        \hline
        \addlinespace
        & Pets 1 & &88.7 & 53.4 & 66.7 & 17.7 & 74.3  \\
        \hline
        \addlinespace
        & Student Transcript Tracking & & 92.3 & 45.1 & 60.5 & 18.6 & 84.9\\
        \hline
        \addlinespace
        & Dog Kennels & & 80.7 & 29.1 & 42.8 & 22.0 & 78.4 \\
        \hline
        \addlinespace
        & Car 1 & & 87.2 & 37.0 & 51.9 & 16.4 & 84.5 \\
        \hline
        \addlinespace
        & Course Teach & & 72.5 & 37.5 & 49.4 & 32.8 & 56.9 \\
        \hline
        \addlinespace
        & World 1 & &75.4 & 10.3 & 18.1 & 21.2 & 76.5  \\
        \hline \hline
    \end{tabular}
\caption{The table presents metrics for the \textbf{token-level prod} method using the Llama 70B model on the Spider dataset using the schema-level approach for each database for a threshold of 0.9. The metrics include Precision \textbf{(P)}, Recall \textbf{(R)}, F1-score \textbf{(F1)}, Isotonic-scaled Brier score (BS-I) and area under the ROC curve (AUC).}
\label{tab:per-schema}
\end{table*}

\begin{table*}[ht]
    \centering
    \fontsize{9}{10}\selectfont
    \addtolength{\tabcolsep}{-0.5em}
    \def\arraystretch{1.0}
    \begin{tabular}{p{0.3cm} p{4.3cm} p{0.7cm} | R{1.1cm} R{1.1cm} R{1.1cm} R{1.3cm} R{1.3cm}}
        \toprule
        \multicolumn{3}{c|}{\textbf{Schema}} & \textbf{P$\uparrow$} & \textbf{R$\uparrow$} & \textbf{F1$\uparrow$} & \textbf{BS-I$\downarrow$} & \textbf{AUC$\uparrow$}\\
        \midrule \midrule
        \addlinespace
        & Thrombosis Prediction && 76.9 & 13.4 & 22.8 & 11.7 & 81.2 \\
        \hline
        \addlinespace
        & Formula 1 && 65.1 & 05.8 & 10.6 & 22.2 & 66.1 \\
        \hline
        \addlinespace
        & California Schools && 74.1 & 14.7 & 24.5 & 13.2 & 83.4\\
        \hline
        \addlinespace
        & Codebase Community && 83.1 & 14.2 & 24.3 & 21.7 & 74.8\\ 
        \hline
        \addlinespace
        &Toxicology&&65.6 & 06.8 & 12.4 & 17.2 & 77.4  \\
        \hline
        \addlinespace
        &Card Games&&54.5 & 01.7 & 3.2 & 17.1 & 74.1  \\
        \hline
        \addlinespace
        &Financial&&52.2 & 05.9 & 10.6 & 17.3 & 76.5  \\
        \hline
        \addlinespace
        &European Football 2&&52.6 & 03.6 & 06.7 & 18.6 & 74.2 \\
        \hline
        \addlinespace
        &Superhero&&75.6 & 11.4 & 19.9 & 22.4 & 75.3  \\        \hline
        \addlinespace
        &Debit Card Specializing&&40.0 & 01.7 & 03.3 & 17.4 & 74.6\\ 
        \hline
        \addlinespace
        &Student Club&&76.5 & 16.9 & 27.7 & 20.9 & 78.4 \\
        \hline \hline
    \end{tabular}
\caption{The table presents metrics for the \textbf{token-level prod} method using the  Llama 70B model on the BIRD dataset using the schema-level approach for each database for a threshold of 0.9. The metrics include Precision \textbf{(P)}, Recall \textbf{(R)}, F1-score \textbf{(F1)}, Isotonic-scaled Brier score (BS-I), area under the ROC curve (AUC), Platt-scaled expected calibration error (ECE-P) and Isotonic-scaled ECE (ECE-I).}
\label{tab:per-schema-level-bird}
\end{table*}

\begin{table*}[ht]
    \centering
    \fontsize{9}{10}\selectfont
    \addtolength{\tabcolsep}{-0.5em}
    \def\arraystretch{1.0}
    \begin{tabular}{p{1.6cm} p{0.3cm} p{0.7cm} | R{1.1cm} R{1.1cm} R{1.1cm} | R{1.1cm} R{1.1cm} R{1.1cm}}
        \toprule
        \multicolumn{3}{c|}{\multirow{2}{*}{\textbf{Threshold}}} & \multicolumn{3}{|c|}{\textbf{Schema-Level}} & \multicolumn{3}{c}{\textbf{Schema-Disjoint}}\\
        \cline{4-9}
        \addlinespace
        & & & \textbf{P$\uparrow$} & \textbf{R$\uparrow$} & \textbf{F1$\uparrow$} & \textbf{P$\uparrow$} & \textbf{R$\uparrow$} & \textbf{F1$\uparrow$}\\
        \midrule \midrule
        \addlinespace
        \multirow{4}{2cm}{\fontsize{9}{11}\selectfont \textbf{Prod}\\(Llama 70B)} & \multirow{4}{*}{$\left\{\rule{0mm}{8mm}\right.$} & 0.9 & 86.1 & 57.3 & 66.9 & 90.4 & 41.0 & 56.5 \\
        && 0.85 & 85.6 & 63.1 & 71.3 & 85.4 & 53.7 & 65.9\\
        && 0.8 & 84.4 & 66.2 & 73.3 & 82.8 & 66.7 & 73.9 \\
        && 0.7 & 83.7 & 74.8 & 78.6 & 80.2 & 70.4 & 75.0 \\
        \hline \hline
    \end{tabular}
\caption{The table presents evaluation metrics for the \textbf{token-level prod} method using the Llama 70B model on the \textbf{Spider} dataset in the schema-level and schema-disjoint approach. The metrics include Precision \textbf{(P)}, Recall \textbf{(R)}, F1-score \textbf{(F1)}.}
\label{tab:prod-spider}
\end{table*}

\begin{table*}[ht]
    \centering
    \fontsize{9}{10}\selectfont
    \addtolength{\tabcolsep}{-0.5em}
    \def\arraystretch{1.0}
    \begin{tabular}{p{1.6cm} p{0.3cm} p{0.7cm} | R{1.1cm} R{1.1cm} R{1.1cm} | R{1.1cm} R{1.1cm} R{1.1cm}}
        \toprule
        \multicolumn{3}{c|}{\multirow{2}{*}{\textbf{Threshold}}} & \multicolumn{3}{|c|}{\textbf{Schema-Level}} & \multicolumn{3}{c}{\textbf{Schema-Disjoint}}\\
        \cline{4-9}
        \addlinespace
        & & & \textbf{P$\uparrow$} & \textbf{R$\uparrow$} & \textbf{F1$\uparrow$} & \textbf{P$\uparrow$} & \textbf{R$\uparrow$} & \textbf{F1$\uparrow$}\\
        \midrule \midrule
        \addlinespace
        \multirow{4}{2cm}{\fontsize{9}{11}\selectfont \textbf{Self-check}\\(Llama 70B)} & \multirow{4}{*}{$\left\{\rule{0mm}{8mm}\right.$} & 0.9 & 85.3 & 54.2 & 63.8 & 88.2 & 32.7 & 47.8 \\
        && 0.85 & 84.3 & 60.5 & 68.6 & 83.4 & 54.8 & 66.2\\
        && 0.8 & 83.6 & 67.5 & 73.6 & 83.5 & 62.9 & 71.8 \\
        && 0.7 & 82.4 & 79.2 & 80.4 & 81.1 & 80.6 & 80.8 \\
        \hline \hline
    \end{tabular}
\caption{The table presents evaluation metrics for the \textbf{self-check bool} method using the Llama 70B model on the \textbf{Spider} dataset in the schema-level and schema-disjoint approach. The metrics include Precision \textbf{(P)}, Recall \textbf{(R)}, F1-score \textbf{(F1)}.}
\label{tab:self-spider}
\end{table*}

\begin{table*}[ht]
    \centering
    \fontsize{9}{10}\selectfont
    \addtolength{\tabcolsep}{-0.5em}
    \def\arraystretch{1.0}
    \begin{tabular}{p{1.6cm} p{0.3cm} p{0.7cm} | R{1.1cm} R{1.1cm} R{1.1cm} | R{1.1cm} R{1.1cm} R{1.1cm}}
        \toprule
        \multicolumn{3}{c|}{\multirow{2}{*}{\textbf{Threshold}}} & \multicolumn{3}{|c|}{\textbf{Schema-Level}} & \multicolumn{3}{c}{\textbf{Schema-Disjoint}}\\
        \cline{4-9}
        \addlinespace
        & & & \textbf{P$\uparrow$} & \textbf{R$\uparrow$} & \textbf{F1$\uparrow$} & \textbf{P$\uparrow$} & \textbf{R$\uparrow$} & \textbf{F1$\uparrow$}\\
        \midrule \midrule
        \addlinespace
        \multirow{4}{2cm}{\fontsize{9}{11}\selectfont \textbf{Prod}\\(Llama 70B)} & \multirow{4}{*}{$\left\{\rule{0mm}{8mm}\right.$} & 0.9 & 65.1 & 08.7 & 15.1 & 72.8 & 02.8 & 05.4 \\
        && 0.85 & 65.4 & 10.1 & 17.0 & 74.2 & 06.7 & 12.2\\
        && 0.8 & 64.0 & 11.1 & 18.5  & 74.6 & 06.9 & 12.6 \\
        && 0.7 & 60.4 & 17.9 & 26.2 & 65.4 & 12.1 & 20.4 \\
        \hline \hline
    \end{tabular}
\caption{The table presents evaluation metrics for the \textbf{token-level prod} method using the Llama 70B model on the \textbf{BIRD} dataset in the schema-level and schema-disjoint approach. The metrics include Precision \textbf{(P)}, Recall \textbf{(R)}, F1-score \textbf{(F1)}.}
\label{tab:prod-bird}
\end{table*}

\begin{table*}[ht]
    \centering
    \fontsize{9}{10}\selectfont
    \addtolength{\tabcolsep}{-0.5em}
    \def\arraystretch{1.0}
    \begin{tabular}{p{1.6cm} p{0.3cm} p{0.7cm} | R{1.1cm} R{1.1cm} R{1.1cm} | R{1.1cm} R{1.1cm} R{1.1cm}}
        \toprule
        \multicolumn{3}{c|}{\multirow{2}{*}{\textbf{Threshold}}} & \multicolumn{3}{|c|}{\textbf{Schema-Level}} & \multicolumn{3}{c}{\textbf{Schema-Disjoint}}\\
        \cline{4-9}
        \addlinespace        
        & & & \textbf{P$\uparrow$} & \textbf{R$\uparrow$} & \textbf{F1$\uparrow$} & \textbf{P$\uparrow$} & \textbf{R$\uparrow$} & \textbf{F1$\uparrow$}\\
        \midrule \midrule
        \addlinespace
        \multirow{4}{2cm}{\fontsize{9}{11}\selectfont \textbf{Self-check}\\(Llama 70B)} & \multirow{4}{*}{$\left\{\rule{0mm}{8mm}\right.$} & 0.9 & 64.5 & 21.7 & 20.3 & 86.3 & 01.8 & 03.5 \\
        && 0.85 & 64.2 & 24.8 & 24.8 & 74.2 & 08.8 & 15.7\\
        && 0.8 &  64.4 & 26.8 & 26.9 & 73.8 & 12.4 & 21.3 \\
        && 0.7 & 58.5 & 33.6 & 33.4  & 72.9 & 17.9 & 28.7 \\
        \hline \hline
    \end{tabular}
\caption{The table presents evaluation metrics for the \textbf{self-check bool} method using the Llama 70B model on the \textbf{BIRD} dataset in the schema-level and schema-disjoint approach. The metrics include Precision \textbf{(P)}, Recall \textbf{(R)}, F1-score \textbf{(F1)}.}
\label{tab:self-bird}
\end{table*}

\section{Reliability plots}
\begin{figure*} 
    \centering
    \vspace{-4mm} 
    \subfloat{\includegraphics[width=\textwidth]{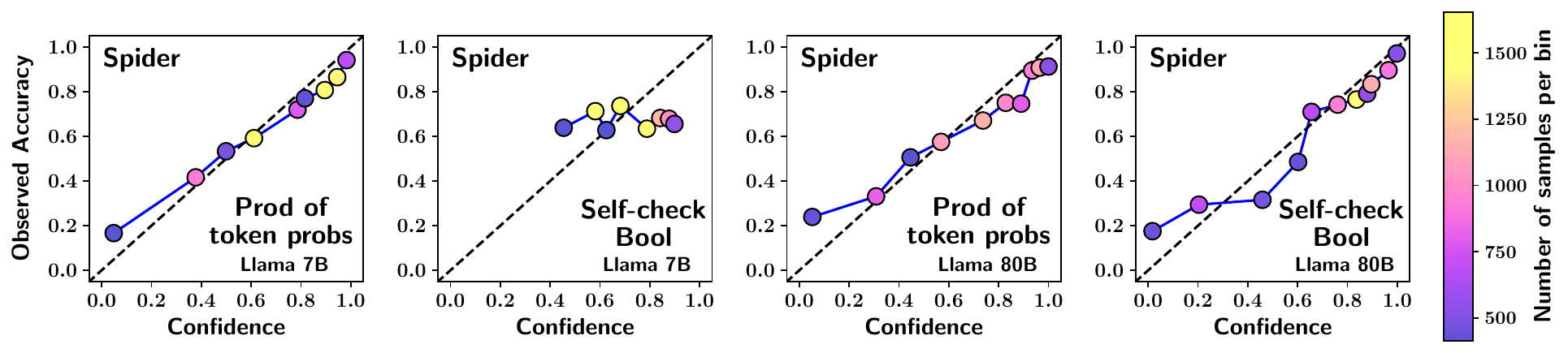}}\\
    \subfloat{\includegraphics[width=\textwidth]{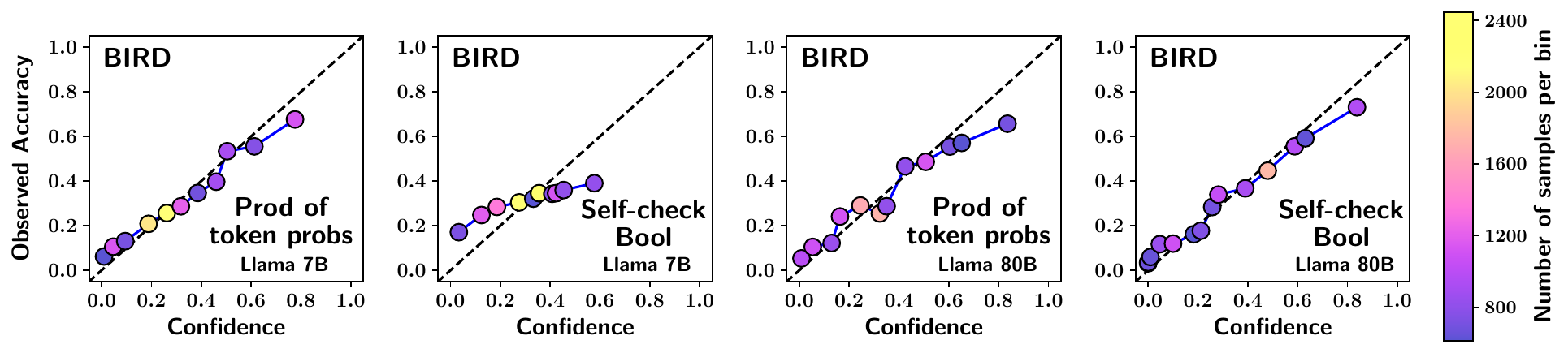}}\vspace{-2mm}
    \caption{The plots have been generated using Monotonic binning in place of Uniform binning  used in ~\ref{fig:plot-1}. The four plots on top have been generated with predictions corresponding to the Spider dataset and four plots below, with the BIRD dataset. A well-calibrated plot aligns closely with the x=y line. Each point is color-coded based on the number of samples in the bin, as indicated by the colorbar on the right.}
    \label{fig:plot-3}
\end{figure*}

\begin{figure*} 
    \centering
    \vspace{-4mm} 
    \subfloat{\includegraphics[width=\textwidth]{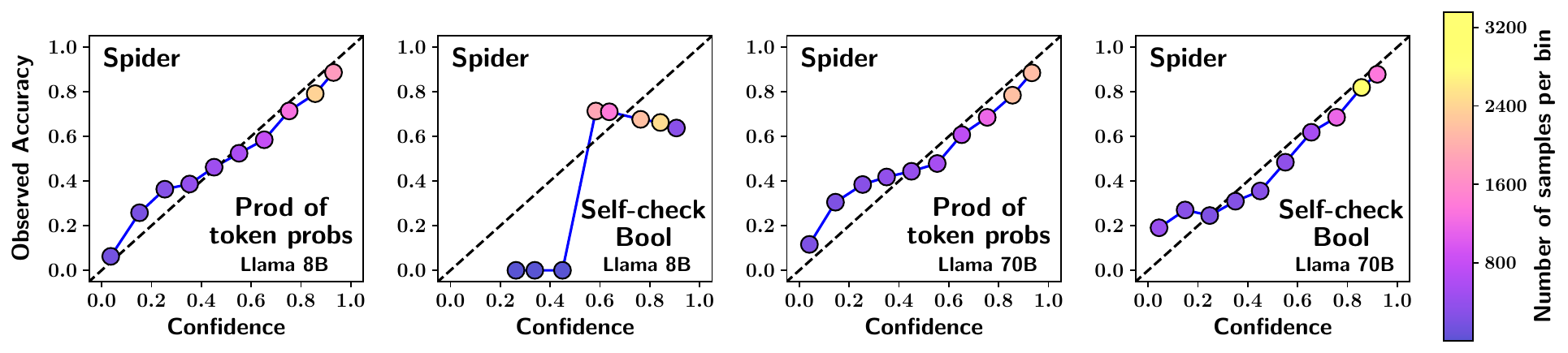}}\\
    \subfloat{\includegraphics[width=\textwidth]{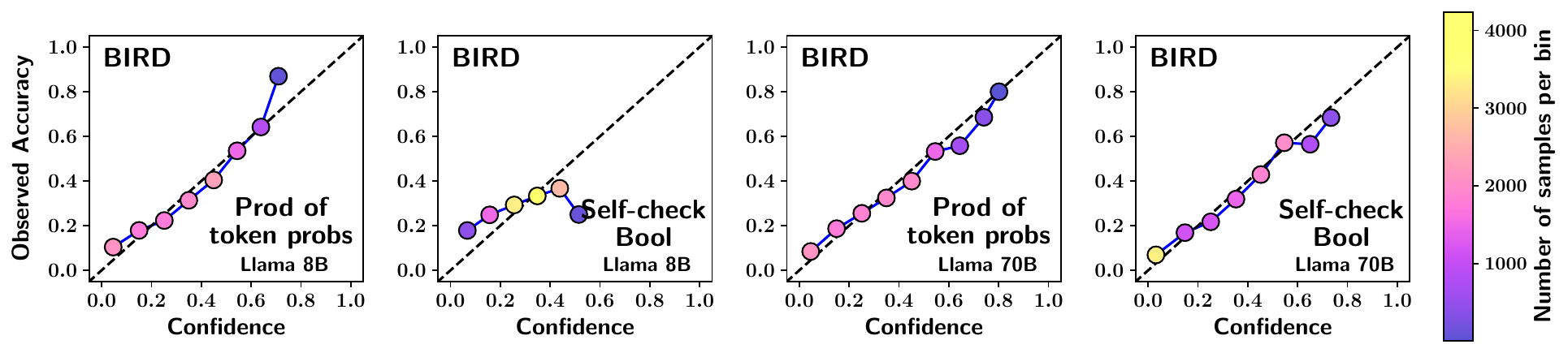}}\vspace{-2mm}
    \caption{The plots have been generated using platt scaling in place of isotonic scaling used in ~\ref{fig:plot-1}. The four plots on top have been generated with predictions corresponding to the Spider dataset and four plots below, with the BIRD dataset. A well-calibrated plot aligns closely with the x=y line. Each point is color-coded based on the number of samples in the bin, as indicated by the colorbar on the right.}
    \label{fig:plot-2}
\end{figure*}

Expanding on figure~\ref{fig:plot-1}, we plot figures~\ref{fig:plot-2} and ~\ref{fig:plot-3} for the other models and methods present in ~\ref{tab:main-table}.

\paragraph{Comparison with Platt scaling}
Table~\ref{tab:isotonic-whole-query} and figure~\ref{fig:plot-2} shows the variation of the evaluation metrics, brier score and expected calibration error, with the two calibration methods, platt scaling and isotonic regression. Note that the AUC and ECE of the raw confidence scores, which are also reported in table~\ref{tab:main-table} are indifferent to calibration.

\paragraph{Comparison with Monotonic binning}
Table~\ref{tab:monotonic-whole-query} and figure~\ref{fig:plot-3} shows the variation of the evaluation metrics, expected calibration error of the raw and calibrated confidence scores, with the two different methods of binning, uniform and monotonic. Note that the AUC and Brier score, which are also reported in table~\ref{tab:main-table} are indifferent to the binning method.

\begin{figure*} 
    \centering
    \vspace{-4mm} 
    \subfloat{\includegraphics[width=\textwidth]{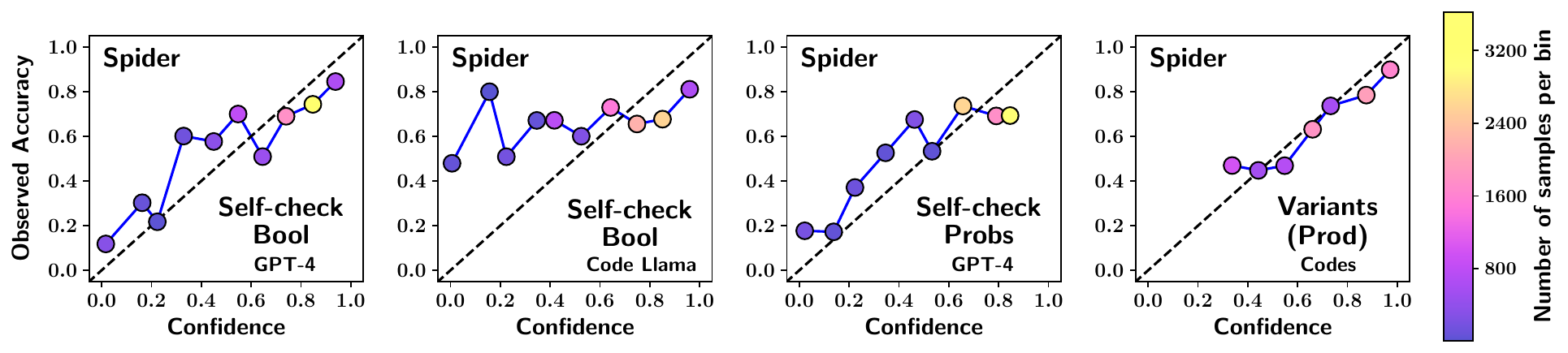}}\\
    \subfloat{\includegraphics[width=\textwidth]{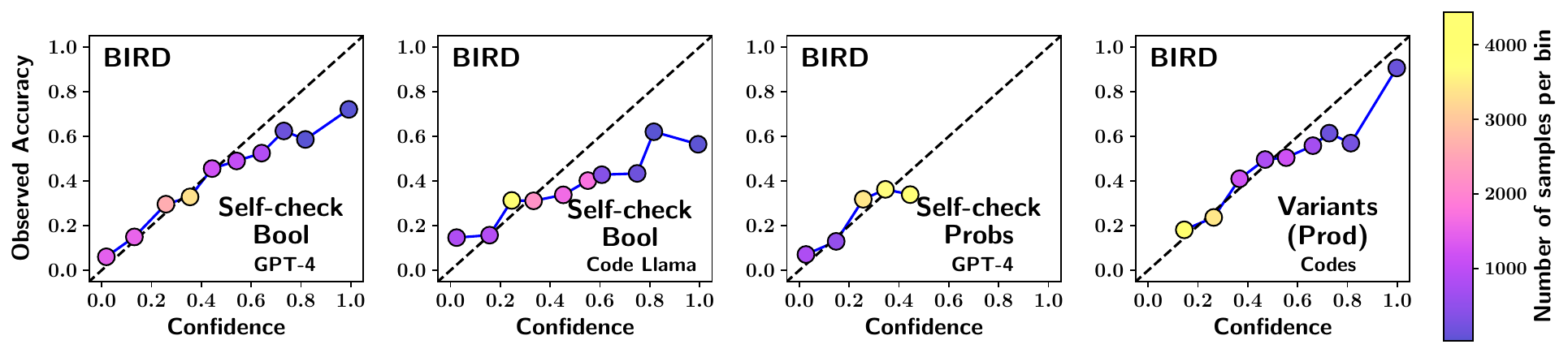}}\vspace{-2mm}
    \caption{The reliability plots continued from ~\ref{fig:plot-1} to illustrate the calibration comparison between the different whole query methods. The four plots on top have been generated with predictions corresponding to the Spider dataset and four plots below, with the BIRD dataset. A well-calibrated plot aligns closely with the x=y line. Each point is color-coded based on the number of samples in the bin, as indicated by the colorbar on the right.}
    \label{fig:plot-4}
\end{figure*}

\begin{figure*} 
    \centering
    \vspace{-4mm} 
    \subfloat{\includegraphics[width=\textwidth]{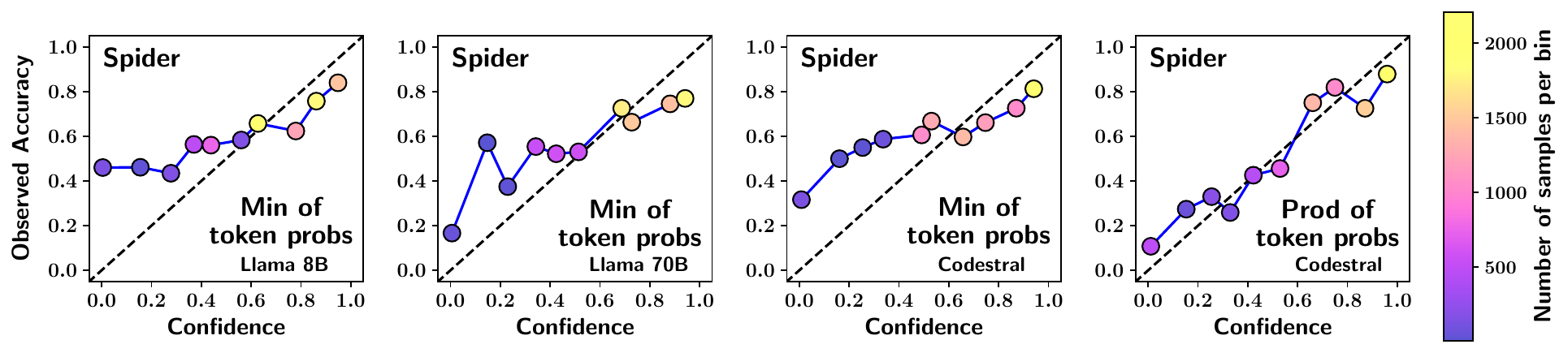}}\\
    \subfloat{\includegraphics[width=\textwidth]{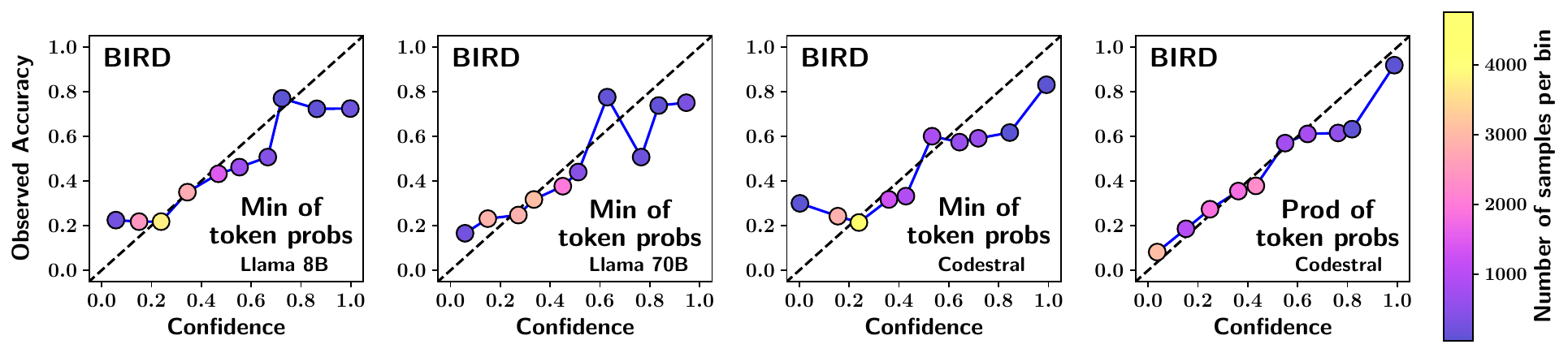}}\vspace{-2mm}
    \caption{The reliability plots continued from ~\ref{fig:plot-1} to illustrate the calibration comparison between the different whole query methods. The four plots on top have been generated with predictions corresponding to the Spider dataset and four plots below, with the BIRD dataset. A well-calibrated plot aligns closely with the x=y line. Each point is color-coded based on the number of samples in the bin, as indicated by the colorbar on the right.}
    \label{fig:plot-5}
\end{figure*}

\begin{table*}[ht]
    \centering
    \def\arraystretch{1.2}
    \begin{tabular}{p{2cm} p{0.4cm} p{0.5cm} | R{2.5cm} R{2.5cm} | R{2.5cm} R{2.5cm}}
        \toprule
        \multicolumn{3}{c|}{\multirow{2}{*}{\textbf{Method}}} & \multicolumn{2}{|c|}{\textbf{Spider}} & \multicolumn{2}{c}{\textbf{BIRD}}\\
        \cline{4-7}
        & &&  \textbf{BS-(P/I)$\downarrow$} & \textbf{ECE-(P/I)$\downarrow$} & \textbf{BS-(P/I)$\downarrow$} & \textbf{ECE-(P/I)$\downarrow$} \\
        \midrule \midrule
        \multirow{13}{0.5cm}{\hspace{-1cm}\begin{sideways}{\fontsize{11}{3}\selectfont\textbf{Isotonic}}\end{sideways}}\hspace{-11.5mm}\ldelim\{{12}{2.5mm}\hspace{2mm}
        \multirow{4}{2cm}{\fontsize{9}{11}\selectfont \textbf{Pooled token-level}\\(Llama 8B)} & \hspace{-0.5cm}\multirow{4}{*}{$\left\{\rule{0mm}{9mm}\right.$} & \hspace{-0.5cm}min & 22.0 $\pm$ 0.01 & 13.0 $\pm$ 0.03 & 20.6 $\pm$ 0.01 & 7.6 $\pm$ 0.02 \\
        && \hspace{-0.5cm}avg & 23.0 $\pm$ 0.01 & 14.1 $\pm$ 0.04 &  19.7 $\pm$ 0.00 & 7.6 $\pm$ 0.02\\
        && \hspace{-0.5cm}\pdt & 18.5 $\pm$ 0.01 & 11.0 $\pm$ 0.04 & 18.9 $\pm$ 0.01 & 7.7 $\pm$ 0.03\\
        && \hspace{-0.5cm}geo & 22.9 $\pm$ 0.02 & 13.6 $\pm$ 0.04 & 20.1 $\pm$ 0.00 & 6.3 $\pm$ 0.02 \\
        \addlinespace
        \hline
        \addlinespace
        \multicolumn{3}{l|}{\parbox{2.8cm}{\fontsize{9}{11}\selectfont\textbf{Self-check Bool}\\(Llama 8B)}} & 23.7 $\pm$ 0.02 & 14.1 $\pm$ 0.04& 22.2 $\pm$ 0.01 & 9.1 $\pm$ 0.03 \\
        \addlinespace
        \hline
        \addlinespace
        \multirow{4}{2cm}{\fontsize{9}{11}\selectfont \textbf{Pooled token-level}\\(Llama 70B)} & \hspace{-0.5cm}\multirow{4}{*}{$\left\{\rule{0mm}{9mm}\right.$} & \hspace{-0.5cm}min & 22.1 $\pm$ 0.02 & 14.1 $\pm$ 0.03 &20.8 $\pm$ 0.01 & 8.1 $\pm$ 0.02  \\
        && \hspace{-0.5cm}avg & 23.4 $\pm$ 0.02 & 14.8 $\pm$ 0.04  & 20.0 $\pm$ 0.01 & 7.2 $\pm$ 0.03 \\
        && \hspace{-0.5cm}\pdt & 18.4 $\pm$ 0.02 & 11.8 $\pm$ 0.05& 19.1 $\pm$ 0.01 & 8.5 $\pm$ 0.03\\
        && \hspace{-0.5cm}geo & 23.5 $\pm$ 0.02 & 14.6 $\pm$ 0.05 & 20.2 $\pm$ 0.01 & 7.0 $\pm$ 0.03\\
        \addlinespace
        \hline
        \addlinespace
        \multicolumn{3}{l|}{\parbox{2.8cm}{\fontsize{9}{11}\selectfont\textbf{Self-check Bool}\\(Llama 70B)}} & 17.5 $\pm$ 0.01 & 11.3 $\pm$ 0.01  & 17.4 $\pm$ 0.01 & 7.2 $\pm$ 0.03 \\
        \addlinespace
        \hline
        \addlinespace
        \multirow{13}{0.5cm}{\hspace{-1cm}\begin{sideways}{\fontsize{11}{11}\selectfont\textbf{Platt}}\end{sideways}}\hspace{-11.5mm}\ldelim\{{12}{2.5mm}\hspace{2mm}
        \multirow{4}{2cm}{\fontsize{9}{11}\selectfont \textbf{Pooled token-level}\\(Llama 8B)} & \hspace{-0.5cm}\multirow{4}{*}{$\left\{\rule{0mm}{9mm}\right.$} & \hspace{-0.5cm}min & 21.3 $\pm$ 0.02 & 12.0 $\pm$ 0.03 & 20.5 $\pm$ 0.01 & 7.3 $\pm$ 0.02\\
        && \hspace{-0.5cm}avg & 22.7 $\pm$ 0.01 & 13.5 $\pm$ 0.04 &  19.5 $\pm$ 0.01 & 6.7 $\pm$ 0.02\\
        && \hspace{-0.5cm}\pdt & 18.1 $\pm$ 0.01 & 10.2 $\pm$ 0.04& 18.9 $\pm$ 0.01 & 8.3 $\pm$ 0.02 \\
        && \hspace{-0.5cm}geo &22.6 $\pm$ 0.02 & 13.2 $\pm$ 0.04& 20.0 $\pm$ 0.01 & 6.4 $\pm$ 0.02  \\
        \addlinespace
        \hline
        \addlinespace
        \multicolumn{3}{l|}{\parbox{2.8cm}{\fontsize{9}{11}\selectfont\textbf{Self-check Bool}\\(Llama 8B)}} & 23.4 $\pm$ 0.02 & 13.1 $\pm$ 0.03&21.7 $\pm$ 0.01 & 6.9 $\pm$ 0.03 \\
        \addlinespace
        \hline
        \addlinespace
        \multirow{4}{2cm}{\fontsize{9}{11}\selectfont \textbf{Pooled token-level}\\(Llama 70B)} & \hspace{-0.5cm}\multirow{4}{*}{$\left\{\rule{0mm}{9mm}\right.$} & \hspace{-0.5cm}min & 21.7 $\pm$ 0.01 & 13.1 $\pm$ 0.04 &20.8 $\pm$ 0.01 & 7.8 $\pm$ 0.02 \\
        && \hspace{-0.5cm}avg &  22.9 $\pm$ 0.02 & 13.6 $\pm$ 0.04  &  19.7 $\pm$ 0.01 & 6.5 $\pm$ 0.03\\
        && \hspace{-0.5cm}\pdt & 18.1 $\pm$ 0.02 & 11.3 $\pm$ 0.05& 18.9 $\pm$ 0.01 & 8.1 $\pm$ 0.03\\
        && \hspace{-0.5cm}geo &23.1 $\pm$ 0.02 & 13.7 $\pm$ 0.04 & 20.1 $\pm$ 0.01 & 6.7 $\pm$ 0.02\\
        \addlinespace
        \hline
        \addlinespace
        \multicolumn{3}{l|}{\parbox{2.8cm}{\fontsize{9}{11}\selectfont\textbf{Self-check Bool}\\(Llama 70B)}} & 17.1 $\pm$ 0.01 & 10.6 $\pm$ 0.01   &17.5 $\pm$ 0.00 & 8.0 $\pm$ 0.02  \\
        \addlinespace
        \hline \hline
    \end{tabular}
\caption{The table compares evaluation metrics across the two calibration methods,  isotonic regression and Platt scaling, for various calibration methods on the Spider and BIRD datasets. The first six rows present isotonic-regression Brier score \textbf{(BS-I)} and isotonic-regression ECE \textbf{(ECE-I)} and the last six rows present Platt-scaled Brier score \textbf{(BS-P)} and Platt-scaled ECE \textbf{(ECE-P)}. Uniform binning is used to calculate ECE-P and ECE-I.}
\label{tab:isotonic-whole-query}
\end{table*}

\begin{table*}[ht]
    \centering
    \def\arraystretch{1.2}
    \begin{tabular}{p{2cm} p{0.4cm} p{0.5cm} | R{2.5cm} R{2.5cm} | R{2.5cm} R{2.5cm}}
        \toprule
        \multicolumn{3}{c|}{\multirow{2}{*}{\textbf{Method}}} & \multicolumn{2}{|c|}{\textbf{Spider}} & \multicolumn{2}{c}{\textbf{BIRD}}\\
        \cline{4-7}
        & & & \textbf{ECE$\downarrow$} & \textbf{ECE-P$\downarrow$} & \textbf{ECE$\downarrow$} & \textbf{ECE-P$\downarrow$} \\
        \midrule \midrule
        \multirow{13}{0.5cm}{\hspace{-1cm}\begin{sideways}{\fontsize{11}{11}\selectfont\textbf{Uniform Binning}}\end{sideways}}\hspace{-11.5mm}\ldelim\{{12}{2.5mm}\hspace{2mm}
        \multirow{4}{2cm}{\fontsize{9}{11}\selectfont \textbf{Pooled token-level}\\(Llama 8B)} & \hspace{-0.5cm}\multirow{4}{*}{$\left\{\rule{0mm}{9mm}\right.$} & \hspace{-0.5cm}min & 67.1 $\pm$ 0.02 & 13.0 $\pm$ 0.03 & 29.1 $\pm$ 0.01 & 7.6 $\pm$ 0.02  \\
        && \hspace{-0.5cm}avg & 6.4 $\pm$ 0.02 & 14.1 $\pm$ 0.04  & 46.0 $\pm$ 0.01 & 7.6 $\pm$ 0.02 \\
        && \hspace{-0.5cm}\pdt & 68.2 $\pm$ 0.02 & 11.0 $\pm$ 0.04&  31.0 $\pm$ 0.01 & 7.7 $\pm$ 0.03  \\   
        && \hspace{-0.5cm}geo & 27.1 $\pm$ 0.03 & 13.6 $\pm$ 0.04& 19.8 $\pm$ 0.01 & 6.3 $\pm$ 0.02 \\
        \addlinespace
        \hline
        \addlinespace
        \multicolumn{3}{l|}{\parbox{2.8cm}{\fontsize{9}{11}\selectfont\textbf{Self-check Bool}\\(Llama 8B)}} & 31.1 $\pm$ 0.03 & 14.1 $\pm$ 0.04 & 7.4 $\pm$ 0.01 & 9.1 $\pm$ 0.03\\
        \addlinespace
        \hline
        \addlinespace
        \multirow{4}{2cm}{\fontsize{9}{11}\selectfont \textbf{Pooled token-level}\\(Llama 70B)} & \hspace{-0.5cm}\multirow{4}{*}{$\left\{\rule{0mm}{9mm}\right.$} & \hspace{-0.5cm}min & 68.2 $\pm$ 0.02 & 14.1 $\pm$ 0.03 &  29.0 $\pm$ 0.01 & 8.1 $\pm$ 0.02 \\
        && \hspace{-0.5cm}avg & 4.9 $\pm$ 0.01 & 14.8 $\pm$ 0.04 & 44.5 $\pm$ 0.01 & 7.2 $\pm$ 0.03 \\
        && \hspace{-0.5cm}\pdt & 68.4 $\pm$ 0.02 & 11.8 $\pm$ 0.05 & 31.0 $\pm$ 0.01 & 8.5 $\pm$ 0.03    \\
        && \hspace{-0.5cm}geo & 32.1 $\pm$ 0.03 & 14.6 $\pm$ 0.05&  17.4 $\pm$ 0.01 & 7.0 $\pm$ 0.03\\
        \addlinespace
        \hline
        \addlinespace
        \multicolumn{3}{l|}{\parbox{2.8cm}{\fontsize{9}{11}\selectfont\textbf{Self-check Bool}\\(Llama 70B)}} & 28.6 $\pm$ 0.02 & 11.3 $\pm$ 0.01 & 7.8 $\pm$ 0.01 & 7.2 $\pm$ 0.03 \\
        \addlinespace
        \hline
        \addlinespace
        \multirow{13}{0.5cm}{\hspace{-1cm}\begin{sideways}{\fontsize{11}{11}\selectfont\textbf{Monotonic Binning}}\end{sideways}}\hspace{-11.5mm}\ldelim\{{12}{2.5mm}\hspace{2mm}
\multirow{4}{2cm}{\fontsize{9}{11}\selectfont \textbf{Pooled token-level}\\(Llama 8B)} & \hspace{-0.5cm}\multirow{4}{*}{$\left\{\rule{0mm}{9mm}\right.$} & \hspace{-0.5cm}min & 67.1 $\pm$ 0.02 & 12.8 $\pm$ 0.03 & 29.1 $\pm$ 0.01 & 7.2 $\pm$ 0.02\\
        && \hspace{-0.5cm}avg & 5.8 $\pm$ 0.02 & 13.5 $\pm$ 0.03 & 46.0 $\pm$ 0.01 & 7.4 $\pm$ 0.02 \\
        && \hspace{-0.5cm}\pdt &68.2 $\pm$ 0.02 & 10.8 $\pm$ 0.04& 31.0 $\pm$ 0.01 & 8.1 $\pm$ 0.02\\
        && \hspace{-0.5cm}geo & 27.1 $\pm$ 0.03 & 13.5 $\pm$ 0.04    & 19.6 $\pm$ 0.01 & 6.3 $\pm$ 0.02\\
        \addlinespace
        \hline
        \addlinespace
        \multicolumn{3}{l|}{\parbox{2.8cm}{\fontsize{9}{11}\selectfont\textbf{Self-check Bool}\\(Llama 8B)}} & 30.6 $\pm$ 0.03 & 14.1 $\pm$ 0.04 &  7.0 $\pm$ 0.01 & 9.0 $\pm$ 0.03 \\
        \addlinespace
        \hline
        \addlinespace
        \multirow{4}{2cm}{\fontsize{9}{11}\selectfont \textbf{Pooled token-level}\\(Llama 70B)} & \hspace{-0.5cm}\multirow{4}{*}{$\left\{\rule{0mm}{9mm}\right.$} & \hspace{-0.5cm}min & 68.2 $\pm$ 0.02 & 14.1 $\pm$ 0.04 & 29.0 $\pm$ 0.01 & 7.9 $\pm$ 0.02\\
        && \hspace{-0.5cm}avg & 4.3 $\pm$ 0.01 & 14.3 $\pm$ 0.04  & 44.5 $\pm$ 0.01 & 7.6 $\pm$ 0.02 \\
        && \hspace{-0.5cm}\pdt & 68.4 $\pm$ 0.02 & 11.8 $\pm$ 0.04 &  31.0 $\pm$ 0.01 & 8.6 $\pm$ 0.03 \\
        && \hspace{-0.5cm}geo &32.1 $\pm$ 0.03 & 14.4 $\pm$ 0.04 & 17.4 $\pm$ 0.01 & 6.9 $\pm$ 0.02\\
        \addlinespace
        \hline
        \addlinespace
        \multicolumn{3}{l|}{\parbox{2.8cm}{\fontsize{9}{11}\selectfont\textbf{Self-check Bool}\\(Llama 70B)}} & 28.6 $\pm$ 0.02 & 11.0 $\pm$ 0.01 &  7.3 $\pm$ 0.01 & 6.7 $\pm$ 0.03 \\
        \addlinespace
        \hline \hline
    \end{tabular}
\caption{The table compares evaluation metrics across the two binning method, Uniform Binning and Monotonic Binning, for various calibration methods on the Spider and BIRD datasets. The first six rows present \textbf{ECE} and Platt-scaled ECE \textbf{(ECE-P)} obtained using Uniform binning and the last six rows present \textbf{ECE} and Platt-scaled ECE \textbf{(ECE-P)} obtained using Monotonic binning.}
\label{tab:monotonic-whole-query}
\end{table*}

\end{document}